\documentclass[prb,twocolumn,showpacs,floatfix,superscriptaddress]{revtex4}
\pdfoutput=1
\usepackage{graphicx}
\usepackage{times}
\usepackage{amsmath,amssymb,bm}
\usepackage{dcolumn}
\begin{document}

\title{Excitations and quasi-one-dimensionality in field-induced nematic and spin density wave states}
\author{Oleg A. Starykh}
\affiliation{Department of Physics and Astronomy, University of Utah, Salt Lake
City, UT 84112}
\author{Leon Balents}
\affiliation{Kavli Institute of Theoretical Physics, University of California, Santa Barbara, Santa Barbara, CA, 93106}
\date{December 2, 2013}
\begin{abstract}
  We study the excitation spectrum and dynamical response functions
  for several quasi-one-dimensional spin systems in magnetic fields
  without dipolar spin order transverse to the field.  This includes
  both nematic phases, which harbor ``hidden'' breaking of
  spin-rotation symmetry about the field and have been argued to occur
  in high fields in certain frustrated chain systems with competing
  ferromagnetic and antiferromagnetic interactions, and spin density
  wave states, in which spin-rotation symmetry is truly unbroken.
  Using bosonization, field theory, and exact results on the
  integrable sine-Gordon model, we establish the collective mode
  structure of these states, and show how they can be distinguished
  experimentally.  
\end{abstract}
\maketitle

\section{Introduction}
\label{sec:introduction}

Much of the research in frustrated quantum magnets has focused on the
elusive quest for magnetically disordered phases with highly entangled
ground states: quantum spin liquids \cite{Balents2010}.  Somewhat intermediate between
these rare beasts and commonplace antiferromagnets are moderately
exotic phases of antiferromagnets in strong magnetic fields which
exhibit no dipolar magnetic order transverse to the field, contrary to
typical spin-flop antiferromagnetic states.  One such state, the Spin
Nematic (SN), has received a particularly high degree of theoretical
attention \cite{Andreev1984,Tsunetsugu2006,Penc2011}.  Argued to occur in some quasi-one-dimensional strongly
frustrated insulators with competing ferromagnetic and
antiferromagnetic interactions \cite{Chubukov1991}, the SN phase has a ``hidden'' order
which breaks spin-rotation symmetry about the magnetic field despite
the lack of transverse spontaneous local moments.  A less celebrated
but competitive state in such systems is the collinear Spin Density
Wave (SDW) \cite{extreme,chen2013}, which develops magnetic order but with spontaneous moments, 
whose magnitude is spatially modulated, entirely along the magnetic field direction.  
Both types of phases are strongly quantum,
i.e. cannot occur in classical models with moments of fixed length at
zero temperature.  The absence of transverse moments in both phases
may lead the two to be confused experimentally, and one of the reasons
for the present study is to clearly define the characteristics that
distinguish them in laboratory measurements.

A spin nematic is usually defined as a state without any spontaneous
dipolar order, i.e. so that in a magnetic field along $z$, $\langle
S_i^+ \rangle =0$, but with quadrupolar order, $\langle S_i^+ S_j^+
\rangle \neq 0$, for nearby sites $i,j$.  Such a nematic breaks the $U(1)$
spin rotation symmetry about the field axis, but in a more non-trivial
way than a usual canted antiferromagnet.  The spin nematics relevant
to this paper are based on the frustrated Heisenberg chain with
ferromagnetic nearest-neighbor coupling and antiferromagnetic
second-neighbor coupling, in a strong magnetic field.  For a region of
parameters, the single magnon excitations with $S^z=\pm1$ of the fully
saturated high field state are {\em bound} into pairs with $S^z=\pm
2$.  Roughly, these latter excitations ``condense'' upon lowering the
field, leading to a spin nematic state \cite{Hikihara2008,sudan2009,zhit}.  Some caution should be
exercised, however, since in one dimension true condensation is not
possible, and spontaneous breaking of rotational symmetry about the
field cannot occur.  A sharp characterization of the one-dimensional
(1d) spin nematic is, rather than nematic order, the presence of a
{\em gap} to $S^z=\pm 1$ excitations.  The 1d SN state may be thought
of more properly as a bose liquid of $S^z=\pm 2$ particles, and hence
has not only power-law nematic order but also power-law density
fluctuations of those bosons \cite{Hikihara2008} (see Sec.~\ref{sec:spin-nematic-chains}).
The latter is just power-law SDW correlations.  Inter-chain couplings
can stabilize {\em either} long-range nematic or SDW order.   One of our
results is that, in fact SDW order is typically more stable, and true
nematic long-range order occurs only in a narrow range of
applied fields very close to the fully saturated magnetization.

More generally, SDW order also occurs in frustrated 1d systems from
other mechanisms, unrelated to magnon pairing and 1d spin nematicity.
Thus we will spend considerable time in this paper discussing the
properties of the SDW.  At the level of order parameter, an SDW state
is described by the expectation value
\begin{equation}
  \label{eq:24}
  \langle S_i^z \rangle = M + {\rm Re} \left[ \Phi e^{i{\bf k}_{\rm
        sdw}\cdot {\bf r}_i} \right]+ \cdots
\end{equation}
where the ellipses represent higher order harmonics that may be
present, or small effects from spin-orbit coupling etc.  SDW states
are are relatively common in itinerant systems with Fermi surface
instabilities \cite{Gruner1994}, but much less so at low temperature in insulating spin
systems, which tend to behave classically and hence 
possess magnetic moments of 
fixed length.  From the point of view of symmetry, the SDW breaks no
global symmetries (time reversal symmetry is broken and the $z$ axis
is already selected by a magnetic field), but instead breaks
translational symmetry.  Consequently, its only low energy mode is
expected to be the pseudo-Goldstone mode of these broken translations,
known as a {\em phason}.  The phason is a purely longitudinal mode, as
it corresponds to the phase of $\Phi$ above and hence a modultion only
of $S^z$.  This is also unusual in the context of insulating magnets,
as the low energy collective modes are usually spin waves, which are
{\em transverse} excitations, associated with small rotations of the
spins away from their ordered axes.  In spin wave theory, indeed,
longitudinal modes are typically expected to be highly damped, and
hence either undefined or hard to observe \cite{Affleck1992,Schulz1996}.  In SDW states, they can
instead control the low energy spectral weight in a scattering
experiment.  The SDW state also has transverse excitations, as we
discuss in Sec.~\ref{sec:rpa}, but these exhibit a spectral gap which
is generally non-zero.  They can be distinguished from the phasons by
their polarization and their location in momentum space.
\begin{figure}[h]
\begin{center}
\includegraphics[scale=0.3]{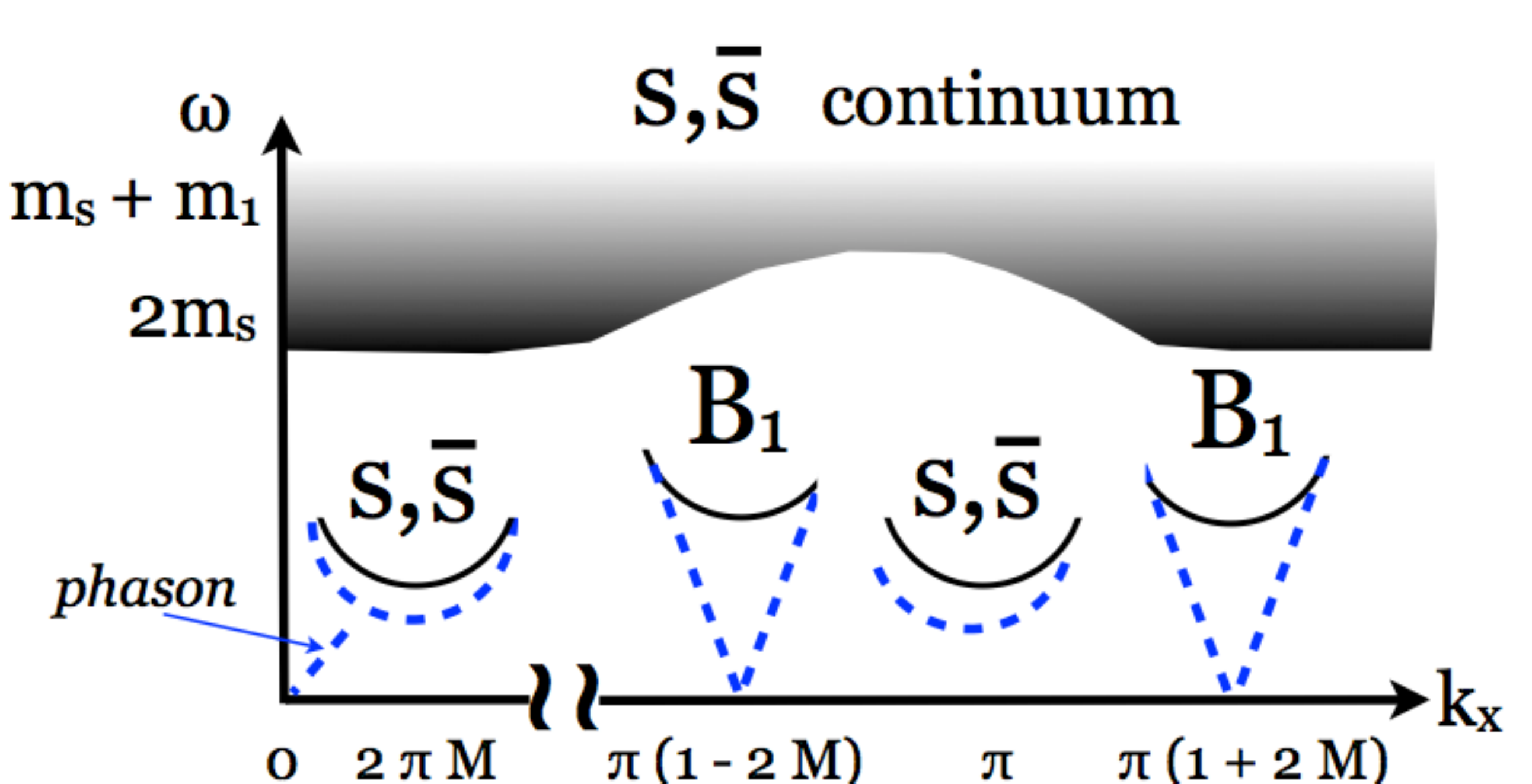}
\caption{Schematic view of the excitation spectrum in the collinear SDW state (see Section~\ref{sec:excit-coll-sdw}),
  i.e. the inelastic structure factor, in momentum (parallel to the
  chain direction defined by strong bonds) and energy space.  Solid
  (black) lines show the results of chain mean field theory,
  i.e. excitations on single chains, and dashed (blue) lines give the
  two-dimensional results corrected for collective inter-chain effects
  (by the RPA approximation).  The symbols $s$ (soliton), $\bar{s}$
  (antisoliton) and $B_1$ (breather) on top of solid lines indicate
  their origin in the excitations of the single chain sine-Gordon
  model.  The excitations shown here at momenta $k_x=\pi (1 \pm 2M)$
  and $k_x=0$ occur in the longitudinal ($S^z$) channel, while those
  at $k_x = \pi$ and $k_x = \pm 2\pi M$ occur in the transverse
  ($S^\pm$) one.  Note that while all excitations are gapped at the
  sine-Gordon level (solid lines), the longitudinal excitations become
  gapless, reflecting the phason mode, once two-dimensional effects
  are included.  The shaded gray area indicates a multi-particle
  continuum composed of solitons, antisolitons $(s,{\bar s})$ and
  breathers $B_1$.  The figure is drawn for the situation $M<1/4$, for
  which $\pi (1 - 2M)$ is larger than $2\pi M$.  For $M>1/4$, the
  corresponding features exchange places in the sketch.}
\label{fig:SG}
\end{center}
\end{figure}

In this paper, we focus primarily on the {\em excitations} of SDW and 2d
spin nematic states.  We show how to use the tools of one dimensional
field theory, combined with the random phase approximation (RPA) and
other methods to obtain both excitations and their contributions to
different components of the dynamical and momentum dependent spin
susceptibilities in a quantitative fashion.  This analysis is greatly
facilitated by the use of copious exact results on the excitations and
correlation functions of the one dimensional sine-Gordon model \cite{essler03,essler04,gnt-book}.   The
results for the excitations of SDW states can also be easily extended
to describe magnetization plateaux, which can be viewed as SDW states
pinned by the commensurate lattice potential \cite{extreme}.  Most of the results for
SDW excitations carry over directly to such plateaux, with the main
modification that the phason develops a small gap due to pinning.

\begin{figure}[h]
\begin{center}
\includegraphics[scale=0.3]{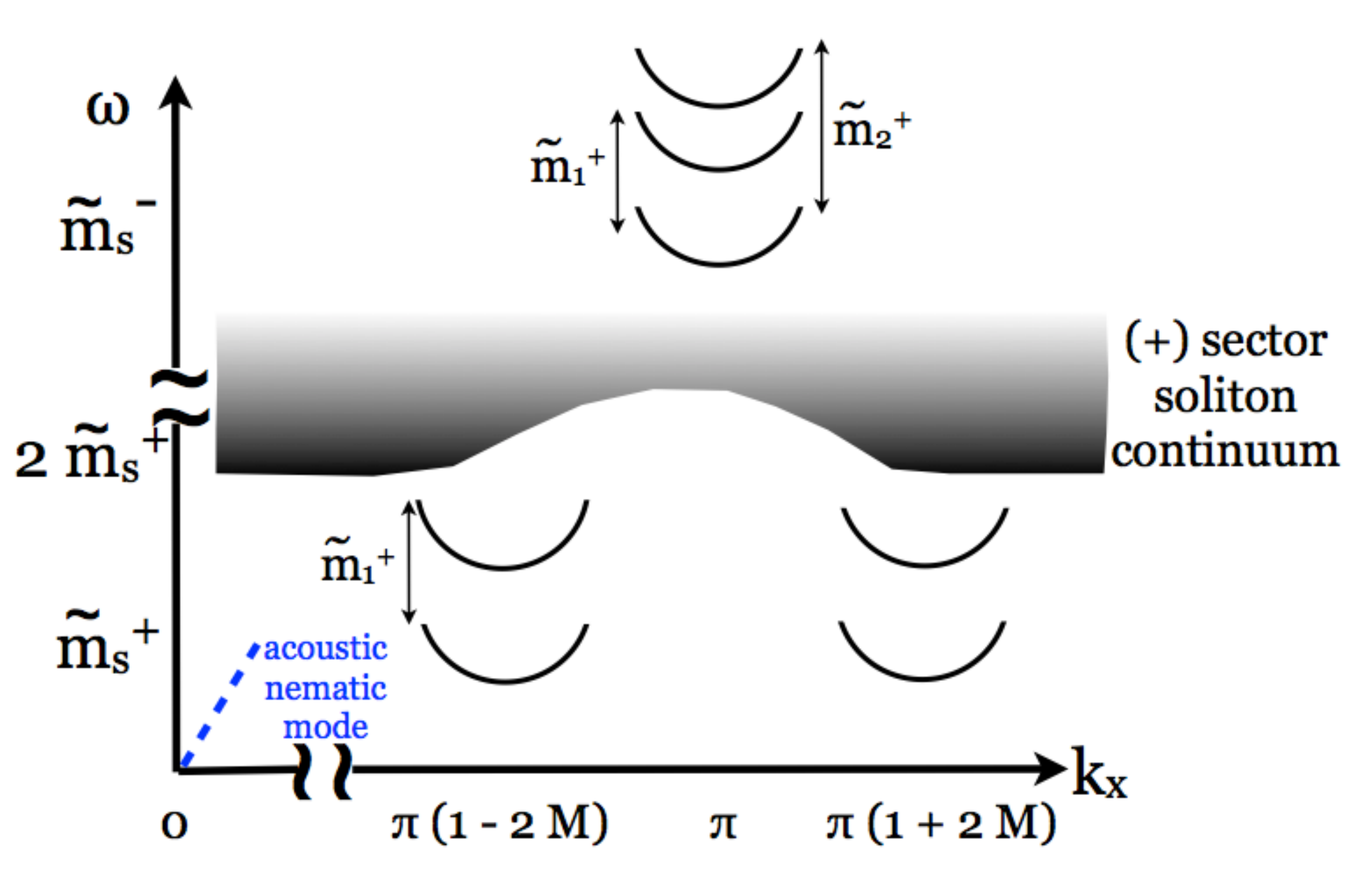}
\caption{ Schematic structure factor, analogous to Fig.~\ref{fig:SG},
  for the two dimensional spin nematic state (see Section~\ref{sec:nematic}).  In contrast to the SDW
  case, only qualitative shifts of the
  excitations away from $k_x=0$ occur, so we draw only solid (black)
  lines there.  Excitations at momenta $k_x=\pi (1 \pm 2M)$ and
  $k_x=0$ are longitudinal, and those at $k_x=\pi$ are
  transverse (a gapped transverse mode at $2\pi M$ is also present, but not shown in the Figure).  
  Note the absence of low energy transverse excitations.
  Indeed, as indicated by the break in vertical scale, excitations at
  $k_x=\pi$ exhibit a much larger gap in the spin nematic case, owing
  to the formation of this gap already at the decoupled chain level.
  The gapless Goldstone mode of the spin nematic, shown as a dashed
  (blue) line, contributes only in
  the vicinity of $k_x=0$.  Vertical axes labels and energy
  separations refer to symbols from the treatment in
  Sec.~\ref{sec:nematic}.  }
\label{fig:nematic}
\end{center}
\end{figure}

In experiment, inelastic neutron
scattering is a powerful way to study the SDW and 2d spin
nematic states, and for convenience we summarize several distinguishing
features identified from our analysis here.   Both states have linearly
dispersion gapless modes: phasons in the SDW case and the Goldstone
modes (``quadrupolar waves'') in the nematic case \cite{Shindou2013,Baryakhtar2013,Smerald2013}.  In the structure
factor, the phason appears with greatest weight at the SDW wavevector,
which is in general incommensurate and away from the zone center and
boundary.  Here it gives a pole contribution whose weight {\em
  diverges} as $1/\omega$ as the energy of the pole approaches zero.
The phason also contributes, although much more weakly, in the vicinity of the
zone center, with a pole whose weight {\em vanishes} as the wave vector
approaches zero.  For the nematic, there is {\em no} divergent gapless
contribution, and the gapless mode appears only at the zone center.
The weights of the zone center contributions, though they both vanish
on approaching $k=0$, differ in the angular dependence of the weight
of the low energy pole.  Another distinction is in the gapped portion
of the spectrum.  In the SDW case, the lowest gapped excitation, which
carries a relatively large spectral weight, occurs usually at
$k_x=\pi$, and occurs in the transverse ($S^\pm$) channel (a caveat
here is that, in the SDW arising out of 1d spin nematic chains, this
is {\em not} the case, and the transverse excitation at $k_x=\pi$ is
pushed to high energy).  In the nematic, the lowest energy gapped
excitations occur instead at the incommensurate value $k_x=\pi(1\pm
2M)$, and excitations at $k_x=\pi$ appear only at much larger
energies.

The rest of the paper is structured as follows.  In
Sec.~\ref{sec:one-dimens-effect}, we introduce bosonization and
one-dimensional effective field theories in a general fashion which
can be applied to both SDW and spin nematic states, in several
different physical contexts.  In Sec.~\ref{sec:excit-coll-sdw} we
derive the excitations and structure factor of the SDW phase, and in
Sec.~\ref{sec:nematic} we do the same for the 2d nematic phase.  We
conclude in Sec.~\ref{sec:discussion} with a Discussion of other ways
to compare SDW and spin nematic phases, and of existing experiments.
Several appendices contain technical details to support the results in
the main text.

\section{One dimensional effective theory}
\label{sec:one-dimens-effect}

In this section, we introduce the standard bosonization description
which applies to many critical one dimensional systems, and establish
notations to be used in the rest of the paper.  A unified formalism of
this type applies to several distinct physical situations, which we
delineate below.  

To justify the bosonization treatment, we will consider a
quasi-one-dimensional geometry, composed of spin chains or ladders,
coupled together by somewhat weaker exchange interactions between these
one dimensional units.  Each unit is characterized by some exchange
scale $J$, presumed the largest in the problem, which sets a temperature
scale $T_{\rm 1d}\sim J$, such that a low energy effective description
of the one dimensional units applies for $T \lesssim T_{\rm 1d}$.
Interactions amongst the one-dimensional units can them be described in
terms of the low energy field theory, i.e. bosonization.  These
interactions, $J' \ll T_{\rm 1d}$, induce ordering with a temperature
$T_{\rm order} \sim J (J'/J)^b \ll T_{\rm 1d}$, where the exponent $b>1$
is in general depedent upon more details of the interactions between and
within the one-dimensional subsystems.  Specific cases will be discussed
below.

\subsection{Bosonization for 1d Bose liquids}
\label{sec:bosonization}

The low energy physics of a great variety of one dimensional spin
systems can be described by bosonization in terms of free scalar
bosonic field theory.   We introduce one such field theory per one
dimensional unit or chain, indexing these units by a discrete variable
$y$.  We  presume $U(1)$ spin rotational symmetry about the $z$ axis,
which allows but does not require a magnetic field along this axis.  

Due to the $U(1)$ symmetry, we may view a spin-1/2 system as a Bose
liquid, mapping for example the $S^z=-1/2$ state to the vacuum, the
$S^z=+1/2$ state to a (hard core) boson, and thereby $S^\pm$ to boson
creation/annihilation operators.  The Bose liquid language has an
advantage in that it allows for a unified view of ordinary
antiferromagnetic spin chains {\sl and} the more exotic one
dimensional nematic (see below).  Therefore we present first the
bosonized form for the theory of a Bose liquid, and then give specific
applications of this to different spin systems.  

For a 1d Bose liquid, the fundamental operators are the density field $n_y(x)$ and
creation/annihilation fields
$\psi_y^\dagger(x),\psi_y^{\vphantom\dagger}(x)$, which are bosonized
(yes, we are bosonizing bosons!)
according to
\begin{eqnarray}
  \label{eq:2}
  n_y(x) &=& \overline{n} + \frac{1}{\beta}\partial_x \varphi_y - A_1
  \sin[\frac{2\pi}{\beta} \varphi_y(x) - k_{\rm sdw} x], \nonumber\\
  \psi_y(x) &=&  A_3 e^{-i \beta \theta_y(x)} + ... 
\end{eqnarray}
Here continuous $x$ runs along the chain, and we have introduced the
slowly-varying ``phase'' fields $\varphi_y(x), \theta_y(x)$ which are
continuous functions of $x$ and time $t$.   The parameter $\beta$
depends upon details of the Bose liquid; it is also often convenient
to introduce the ``compactification radius'' $R = \beta/(2\pi)$.
$\beta$, or equivalently $R$,
determines the long-distance behavior of the 1d correlation
functions.   The modulation wavevector $k_{\rm sdw}$ is that of an incipient Bose
solid at the average Bose density $\overline{n}$, which is $k_{\rm sdw} =
2\pi\overline{n}$.  It is sometimes convenient to define the ``charge
density wave'' order parameter for these bosons,
\begin{equation}
  \label{eq:10a}
  \Phi_y(x) = e^{-i\frac{2\pi}{\beta} \varphi_y},
\end{equation}
so that
\begin{equation}
  \label{eq:11}
   n_y(x) = \overline{n} + \frac{1}{\beta}\partial_x \varphi_y  -\frac{iA_1}{2}\left(
   \Phi_y(x) e^{i k_{\rm sdw} x} - {\rm h.c.}\right).
\end{equation}
For spin systems, $\Phi_y$ becomes the spin density wave order
parameter.  To keep the presentation symmetric, we also define the
``superfluid'' or XY order parameter $\Psi_y = e^{-i \beta \theta_y}$,
so that
\begin{equation}
  \label{eq:13}
  \psi_y(x) = A_3 \Psi_y(x).
\end{equation}

The conjugate fields $\varphi(x), \theta(x)$ obey the commutation relation 
\begin{equation} 
[\theta_y(x), \varphi_{y'}(x')] = - i \Theta(x-x') \delta_{yy'}.
\end{equation}
where $\Theta(x)$ is the Heavyside step-function.  Their dynamics is
described by the free field Hamiltonian
\begin{equation}
H_0 = \sum_y \int dx \frac{v}{2} \{ (\partial_x \theta_y)^2 + (\partial_x \varphi_y)^2\}.
\label{eq:H0}
\end{equation}
This describes a single bosonic mode for each $y$: a central charge
$c=1$ conformal field theory, also known as a Luther-Emery liquid or $c=1$
Luttinger liquid.  The Hamiltonian contains a single parameter $v$,
which gives the velocity of excitations which propagate
relativistically, and which again depends upon microscopic details.

Such a Luttinger liquid is characterized by algebraic correlations,
which are simply obtained from the above free field theory, the most
prominent of which are
\begin{eqnarray}
\label{eq:Sz}
\langle n_y(x) n_y(0) \rangle_c &=& \frac{1}{2} A_1^2 \cos[k_{\rm sdw} x] ~|x|^{-2\Delta_z} , \\
\langle \psi^{\vphantom\dagger}_y(x) \psi^{\dagger}_y(0) \rangle &=& A_3^2 ~|x|^{-2\Delta_\perp} .
\label{eq:Sxy}
\end{eqnarray}
Their power-law decay is controlled by the scaling dimensions
$\Delta_z = \pi/\beta^2 = 1/(4\pi R^2)$ and $\Delta_\perp =
\beta^2/(4\pi)=\pi R^2$.   Here we gave only the leading terms in \eqref{eq:Sz}
and \eqref{eq:Sxy}, omitting corrections which decay faster with distance.

For the case of many spin chains, including the XXZ chain in a field
along $z$, we can simply apply the above bosonization rules taking
\begin{eqnarray}
  \label{eq:1}
  S_y^z(x) & = & \frac{1}{2} - n_y(x), \\
  S_y^+(x) & = & (-1)^x \psi_y(x).
\end{eqnarray}
In that case, $\overline{n} = 1/2 - M$, where $M$ is the uniform
magnetization, and hence $k_{\rm sdw} = \pi - 2\pi M$. 
For the isotropic Heisenberg chain, $2\pi R^2$ monotonically decreases
from $1$ at zero magnetization ($M=0$) to $1/2$ at the full saturation
$M=1/2$. This shows that in the presence of external magnetic field
transverse spin fluctuations are more relevant (decay slower) than the
longitudinal ones, $\Delta_\perp \leq \Delta_z$ for $0 < M \leq
1/2$. At the same time the wave vector of longitudinal spin
fluctuations shifts with magnetization continuously, as $k_{\rm sdw} = \pi(1 -
2 M)$, toward the Brillouin zone center, while that of the transverse
fluctuations, $k_\perp = \pi$, remains fixed at the Brillouin zone
boundary.

As discussed above, two-dimensional order appears as a result of
residual inter-chain interactions $J'$ which are described by a
perturbing Hamiltonian $H'$.  To understand under which conditions
SDW can emerge from $H'$, it is instructive to start by considering the
simplest case of non-frustrated inter-chain coupling
\begin{eqnarray}
\label{eq:Hp-non-fr}
&&H'_{\rm non-fr} = J' \sum_{x,y} {\bf S}_y(x) \cdot {\bf S}_{y+1}(x) \to \\
&&\to \sum_y \int dx~  \gamma_{\rm sdw} \cos[2\pi(\varphi_y -
\varphi_{y+1})/\beta] \nonumber \\
&& + \gamma_{\rm xy} \cos[\beta(\theta_y - \theta_{y+1})] . \nonumber
\end{eqnarray}
Here we rewrote the first line in an appropriate low-energy form with
the help of the representation \eqref{eq:1}, and we defined continuum
inter-chain coupling constants $\gamma_{\rm sdw} =
J' A_1^2$ and $\gamma_{\rm xy} = J' A_3^2$, which are of the same
order. Since the fields on different chains are not correlated with
each other at leading order \eqref{eq:H0}, the scaling dimension $D$
of the SDW (cone) term in \eqref{eq:Hp-non-fr} is simply $D_{\rm sdw}
= 2 \Delta_z (D_{\rm xy} = 2 \Delta_\perp)$.  Since in the case of
isotropic Heisenberg chains $\Delta_\perp \leq \Delta_z$ for all $0 <
M \leq 1/2$, as argued above, the second term in the above equation
becomes parametrically stronger than the first under the
renormalization group (RG) flow.  As a result, the interchain
interaction \eqref{eq:Hp-non-fr} reduces to the {\em xy} term which
implies two-dimensional order, via spontaneous $U(1)$ symmetry
breaking, in the plane perpendicular to the external magnetic field.
This is a familiar canted antiferromagnet, or spin-flop
two sublattice ordered state.  Note that $\langle S^z_y(x)\rangle$ is completely
uniform in this phase.  

The absence of an SDW phase noted here clearly follows from the
condition $D_{\rm sdw}> D_{\rm xy}$.  We observe that this may break
down in three ways.  First, for spin chains other than the simple
Heisenberg one, the inequality $\Delta_\perp < \Delta_z$ may be
violated in favor of the opposite situation.  Second, for yet more
exotic spin chains (or ladders), the relation between spin operators
and those of the effective Bose gas may differ from that in
Eqs.~\eqref{eq:1}.  Finally, third, the interactions
between chains may differ from those in Eq.~\eqref{eq:Hp-non-fr}.  We
will encounter all these situations below.  

\subsection{Physical realizations}
\label{sec:phys-real}

We now consider three different microscopic lattice models that lead
to dominant SDW interactions. These models represent physically
different ways of achieving the inequality $D_{\rm sdw} \leq D_{\rm
  xy}$.  In general, the models we consider have, in their bosonized
continuum limits, a Hamiltonian of the form $H=H_0 + H'$, with $H_0$
describing decoupled chains as in Eq.~\eqref{eq:H0}, and the
inter-chain coupling of the form
\begin{eqnarray}
  \label{eq:12}
  H' & = & \sum_y \int \! dx\, \Big\{ \tfrac{1}{2}\gamma_{\rm sdw} \left(\Phi_y^\dagger
  \Phi_{y+1}^{\vphantom\dagger} + \Phi_{y+1}^\dagger \Phi_y ^{\vphantom\dagger}\right) \nonumber \\
& & + \tfrac{1}{2}\gamma_{\rm xy} \left(\Psi_y^\dagger
  \Psi_{y+1}^{\vphantom\dagger} + \Psi_{y+1}^\dagger \Psi_y ^{\vphantom\dagger}\right) \nonumber \\
& & +\tfrac{1}{2} \gamma'_{\rm xy} \left(\Psi_y^\dagger
  i\partial_x\Psi_{y+1} ^{\vphantom\dagger}- \Psi_{y+1}^\dagger i\partial_x\Psi_y ^{\vphantom\dagger}\right) .
\end{eqnarray}
The different models are distinguished by the values of the couplings
$\gamma_{\rm sdw}, \gamma_{\rm xy}, \gamma'_{\rm xy}$ {\em and} by the
value of the chain interaction parameter $\beta$.  

\begin{table}
\centering 
\begin{tabular}{p{2.5cm}|c|c|c}
{\bf model} & $\gamma_{\rm sdw}$ & $\gamma_{\rm xy}$ & $\gamma'_{\rm
  xy}$  \\
\hline
\hline 
Ising chains & $\tfrac{1}{2}J' \delta A_1^2$ & $J' A_3^2$ & 0 \\ \hline
nematic chains & $\tfrac{1}{2}J' A_1^2$ & $\sim (J')^2/J $ & $0$ \\ \hline
triangular lattice & $J' A_1^2 \sin (\pi M) $ & $0$ & $J'
A_3^2/2$
\end{tabular}
\label{tab:params}
\caption{Parameters describing three different physical
  realizations of the quasi-one-dimensional SDW state.}
\end{table}

\begin{figure}[h]
\begin{center}
\includegraphics[scale=0.2]{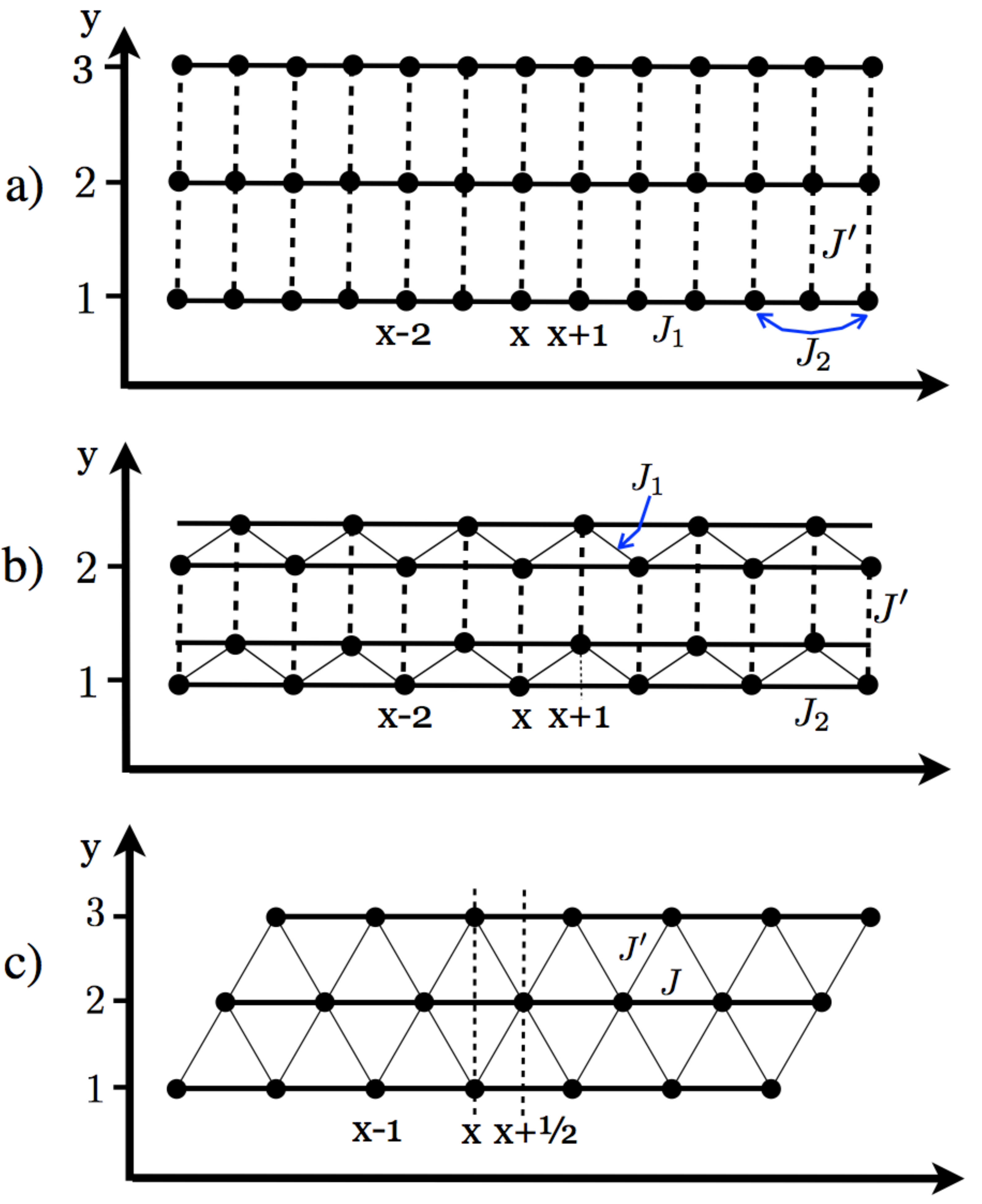}
\caption{ Lattice geometries considered in the paper.  (a) Rectangular
  geometry, relevant for Ising-like coupled chains, discussed in
  Sec.~\ref{sec:ising-anisotropy}, and also for nematic chains,
  considered in Sec.~\ref{sec:spin-nematic-chains}. In the latter case
  $J_1 < 0$ and $J_2 > 0$.  (b) Equivalent representation of coupled
  nematic chains as a system of coupled zig-zag ladders.  (c)
  Spatially anisotropic triangular lattice, discussed in
  Section~\ref{sec:spat-anis-triang}.  }
\label{fig:geometry}
\end{center}
\end{figure}

\subsubsection{Ising anisotropy}
\label{sec:ising-anisotropy}

The most straightforward route to $D_{\rm sdw} \leq D_{\rm xy}$ is
provided by arranging $\Delta_{\perp}>\Delta_{z}$.  This occurs by keeping the
same unfrustrated rectangular arrangement of spin-1/2 chains discussed
above, but replacing the Heisenberg chains with XXZ ones with Ising anisotropy,
\begin{eqnarray}
H_{\rm Ising}&  = &  J \sum_{x,y} (S^x_{x,y} S^x_{x+1,y} + S^y_{x,y}
S^y_{x+1,y} + \delta S^z_{x,y} S^z_{x+1,y}) , \nonumber \\
&  + &  J' \sum_{x,y} (S^x_{x,y} S^x_{x,y+1} + S^y_{x,y}
S^y_{x,y+1} + \delta S^z_{x,y} S^z_{x,y+1}), \nonumber \\
&&
\label{eq:ising}
\end{eqnarray}
where $\delta > 1$ parameterizes Ising anisotropy, and for simplicity
we have taken the same anisotropy in the inter-chain coupling $J'$,
though this is not very important.  In zero magnetic field, even in
the absence of inter-chain coupling, such a chain orders spontaneously
(at zero temperature, $T=0$) into one of the two N\'eel states, with
spins ordered along the easy Ising ($z$) axis.  The the non-frustrated
interchain exchange $J'$ then immediately selects the staggered
arrangement of N\'eel order of adjacent chains, further stabilizing
the antiferromagnet for low but non-zero temperature.  

However, a sufficiently strong magnetic field, applied along the $z$
axis, breaks the gap, driving the XXZ chains into gapless Luttinger
liquid state again \cite{Okunishi2007}. For small $J'$, the problem
can then be treated by bosonization and has the general form found
above in Eq.~\eqref{eq:12}, with $\gamma_{\rm sdw} = J'\delta
A_1^2/2$, $\gamma_{\rm xy} = J' A_3^2$, and $\gamma'_{\rm xy}=0$.
More importantly, the Ising anisotropy increases $\beta$ relative to
the Heisenberg chain.  Indeed, it
turns out that the critical indices of this state (parametrized in
\cite{Okunishi2007} by $\eta$ instead of our $R$) do have the desired
property that $\Delta_z < \Delta_\perp$ for $M$ in the finite range $0
< M \leq M_c(\delta)$. The critical magnetization $M_c(\delta)$,
separating $\Delta_z < \Delta_\perp$ and $\Delta_z > \Delta_\perp$
regimes, increases with increasing anisotropy $\delta >1$.

It is clear that interchain interaction then
stabilizes the two-dimensional SDW state in (approximately) the same
magnetization interval $0 < M \leq M_c(\delta)$ because here $\Delta_z
< \Delta_\perp$ immediately implies $D_{\rm sdw} < D_{\rm cone}$.  The
exact value of the critical magnetization separating the {\em
  two-dimensional} SDW and cone states (with non-zero $J'$) depends on
many details and is not rigorously known.  A reasonable estimate can
be made by the chain mean field theory (CMFT), using the precise forms of the
longitudinal and transverse spin susceptibilities as well as small (of
the order $J'/J \ll 1$) corrections to magnetization $M$ caused by the
interchain exchange $J'$. We disregard all these complications in
order not to overload the discussion.  

It appears that spin-1/2 antiferromagnet BaCo$_2$V$_2$O$_8$ realizes exactly this situation \cite{Kimura2008}.
Static SDW order has been observed in several neutron and sound-attenuation studies (refs).

\subsubsection{Spin-nematic chains}
\label{sec:spin-nematic-chains}

A second route to the collinear SDW is to suppress the leading {\em xy}
instability altogether, by driving the individual spin chain into a
completely different {\sl phase}.  This occurs in the model derived
from LiVCuO$_4$, in which the one-dimensional chains are not XXZ like
but instead incorporate   {\em ferromagnetic} nearest-neighbor
exchange $J_1 < 0$ and antiferromagnetic next-nearest exchange $J_2 >
0$ \cite{Enderle2005,Nishimoto2012}.  Such $J_1 - J_2$ chains (which can also be ``folded'' into
zig-zag ladders) have distinct behavior which is {\em not} captured by
Eqs.~\eqref{eq:1}.  

Extensive research into this interesting chain geometry, dating back to
1991 \cite{Chubukov1991}, has found that the spectrum of the fully
magnetized chain contains, in addition to usual single magnon states,
tightly bound magnon {\em pairs} (in fact, three- and four-magnon
complexes exists in some parameter range as well \cite{Hikihara2008,sudan2009}). Importanly, these
two-magnon pairs lie below the two-magnon continuum. As the magnetic field
is reduced to the critical $h_{\rm sat}$ one, the gap for the
two-magnon states vanishes {\em while the single magnon gap remains
  non-zero}.  For $h<h_{\rm sat}$, therefore, one obtains not a Bose liquid
of single magnons (which is the physical content of
Eqs.~\eqref{eq:1}), but rather a Bose liquid of magnon pairs\cite{Hikihara2008,sato}.  In such
a liquid, Eqs.~\eqref{eq:1} is replaced by
\begin{eqnarray}
  \label{eq:3}
    S_y^z(x) & \sim & \frac{1}{2} - 2n_y(x), \nonumber\\
  S_y^+(x)S_y^+(x+1) & \sim & \psi_y(x).
\end{eqnarray}
where now $\psi_y(x)$ annihilates a magnon pair, and $n_y(x)$ counts
the magnon pairs.  The appearance of the operator quadratic in $S^+_y$
above indicates the existence of critical ``spin nematic''
correlations.  Since a gap for single magnons (single spin flips)
remains, the low energy projection of the single spin-flip operator
vanishes
\begin{equation}
  \label{eq:4}
  S_y^\pm (x) \sim``0".
\end{equation}
For a single $J_1-J_2$ chain, this is still a Luttinger liquid state,
but simple XY correlations decay exponentially instead of as a power law.  The density correlations
in this Bose liquid remain critical, and hence from Eq.~\eqref{eq:3}
so do those of $S^z_y(x)$.  

With this understanding, we see that even simple unfrustrated $J'$
exchange interactions coupling the $J_1-J_2$ chains are ``projected''
onto dominantly Ising $S^z_y$ interactions, which strongly favor an
SDW ground state.  Specifically, we have again the form in
Eq.~\eqref{eq:12}, but with $\gamma_{\rm sdw} \sim J' A_1^2 \gg \gamma_{\rm xy} \sim (J')^2/J$
and $\gamma'_{\rm xy} = 0$.  The strong
suppression of all single spin-flip operators suggests that, unlike in
the previous case, the SDW state extends up to very close to the
saturation value $M \sim 1/2$.  

Unusual functional form of  $\gamma_{\rm xy} \sim (J')^2/J$ 
is due to the fact that it describes coupling of the nematic fields $\psi_y(x)$ of different chains.
Such a coupling, involving {\em four} spin operators, see \eqref{eq:3},  
is simply absent in the lattice model. It is, however, generated by quantum fluctuations in
second order in the inter-chain exchange, which explains its peculiar form
(the proportionality constant is non-trivial \cite{sato} and not
determined here).  We will see that this can stabilize a true 2d SN
near the saturation field -- see Sec.~\ref{sec:comp-betw-2d}.  But
away from a narrow region near saturation, the SDW state indeed
dominates as na\"ively expected.

\subsubsection{Spatially anisotropic triangular lattice antiferromagnet}
\label{sec:spat-anis-triang}

In the above two examples, we modified the interactions on the
individual chains from the Heisenberg type.  A third way to stabilize
the SDW phase is to retain the simple nearest-neighbor Heisenberg form
for the chain Hamiltonian, but modify explicitly the interactions
between chains in a manner that {\em frustrates} the competing XY
order.  This occurs naturally for the situation of a spatially anisotropic
triangular lattice \cite{Starykh2007,extreme}.  In this case, each spin is coupled symmetrically
to two neighbors on adjacent chains, which frustrates the inter-chain
interactions.  Specifically, the interchain coupling reads
\begin{eqnarray}
H'_{\rm frust} & & =  \\
&&  J' \sum_{x,y} {\bf S}_y(x) \cdot ({\bf
  S}_{y+1}(x-1/2) + {\bf S}_{y+1}(x+1/2)).\nonumber
\label{eq:Hp-frust}
\end{eqnarray}
Note that this Hamiltonian is written in a cartesian basis in which spins on, say, odd chains are located at the integer positions $x$
while those on the even chains are at the half-integer locations
$x+1/2$.   Bosonization of \eqref{eq:Hp-frust} gives again the form of
Eq.~\eqref{eq:12}, but with $\gamma_{\rm xy}=0$ due to frustration.
The other two interactions are $\gamma_{\rm sdw}= J' A_1^2 \sin (\pi
M)$ and $\gamma'_{\rm xy} = J' A_3^2 /2$. 

The SDW term retains its form but its coupling constant reflects
frustration as well, $\gamma_{\rm sdw} \sim \sin[\pi M] \to 0$
for $M\to 0$.  The SDW coupling resists the appearance of the
derivative which occurs for the XY term, as a result of the shift of
the longitudinal wave vector $k_z = \pi (1 - 2M)$ from its
commensurate value $\pi$ for finite $M\neq 0$. It is this shift that
makes SDW interaction more relevant than the XY one.  While the SDW
scaling dimension remains $D_{\rm sdw} = 2 \Delta_z$, that of the XY
interaction increases to $D_{\rm xy} = 1 + 2 \Delta_\perp$. The
addition of $1$ reflects the derivative in the $\gamma'_{\rm xy}$ term
of Eq.~\eqref{eq:12}.

Since $D_{\rm sdw} = 2 \Delta_z < D_{\rm xy} = 1 + 2 \Delta_\perp$
in a rather wide range of magnetization, approximately for $0 < M \leq
0.3$, interchain frustration stabilizes collinear SDW order \cite{extreme}.

\section{Excitations of collinear SDW state}
\label{sec:excit-coll-sdw}

In this section, we discuss the excitation spectrum of the collinear
SDW state, and its manifestation in the magnetic structure factor (or
wavevector dependent spin susceptibility).  The magnetic excitations
are collective modes, strongly influenced by symmetry.  In an applied
magnetic field, the only symmetries of the Hamiltonian are $U(1)$
rotation symmetry about the field, and the space group symmetries of
the lattice.  Notably, the collinear SDW state preserves the former
$U(1)$ symmetry, and in the absence of broken continuous symmetry,
lacks a Goldstone mode.  Thus there are no acoustic transverse spin
waves.  Instead, we expect gapped transverse excitations.  Given the
highly quantum nature of the SDW phase in the quasi-1d, $S=1/2$
situation discussed here, there is in fact no {\em a priori} reason
these excitations may be treated semiclassically in the traditional
spin wave fashion.  Instead, in the following, we will obtain the gapped
excitations from a purely quantum treatment based on knowledge of
the integrable 1d sine-Gordon model.

The collinear SDW {\em does}, however, break translation symmetry, and
in particular exhibits {\em incommensurate} order (see Eq.~\eqref{eq:24}).
Although translational symmetry is discrete, in cases of
incommensurate order it is known to behave in some respects like a
continuous symmetry and consequently the collinear SDW state supports
a {\em phason} mode, which is the ``pseudo-Goldstone'' mode of broken
translation symmetry.  Physically this mode -- which is acoustic --
appears because of the vanishing energy cost for uniformly ``sliding''
the incommensurate density wave.  In the bosonization framework, the
elevation of the discrete lattice translation symmetry to an
effectively continuous one appears in an emergent continuous symmetry
of Eq.~\eqref{eq:Hp-non-fr}: invariance under $\varphi_y(x) \to
\varphi_y(x) + \varphi^{(0)}$.  While it is well-known in SDW-ordered
metals, the phason excitation is perhaps less
familiar in magnetically ordered insulators. We now turn to the
detailed exposition of the excitation spectrum, including both phason
and gapped modes. For simplicity, we focus here on zero-temperature
($T=0$) properties and apply CMFT to the problem.  An alternative
derivation of the phason dispersion, based on the Ginzburg-Landau (GL)
action, is sketched in Appendix~\ref{ap:phason}.

\subsection{Single chain excitations}
\label{sec:single-chain-excit}

In this subsection, we present the chain mean field theory which
approximates the problem of the 2d system by a self-consistent set of
independent chains, specifically 1+1d sine-Gordon models.  We describe
the gapped excitations occuring within individual such chains.  The
effects of two-dimensionality on the spectrum, and especially the
emergence of the low energy phason mode, is discussed in the following
subsection. 

\subsubsection{Chain mean field theory}
\label{sec:chain-mean-field-1}

Focusing on the SDW state, we drop the $\gamma_{\rm xy}$ and
$\gamma'_{\rm xy}$ terms in Eq.~\eqref{eq:12}, and make the mean field
replacement $H' \rightarrow H'_{\rm MF}$ (neglecting a constant), with
\begin{eqnarray}
  \label{eq:15}
  H'_{\rm MF} & \rightarrow &  = \sum_y \int \! dx\, \tfrac{1}{2}\gamma_{\rm sdw}
  \big( \langle \Phi_y^\dagger\rangle
  \Phi_{y+1}^{\vphantom\dagger} + \Phi_y^\dagger
  \langle \Phi_{y+1}^{\vphantom\dagger}\rangle \nonumber \\
  && + \langle\Phi_{y+1}^\dagger\rangle \Phi_y ^{\vphantom\dagger}+
  \Phi_{y+1}^\dagger \langle \Phi_y ^{\vphantom\dagger}\rangle \big) -
  {\rm const.} .
\end{eqnarray}
With the ansatz
\begin{equation}
  \label{eq:14}
  \langle \Phi_y \rangle  = \overline\Phi (-1)^y,
\end{equation}
we then obtain
\begin{eqnarray}
  \label{eq:16}
  H'_{\rm MF} & = & - \gamma_{\rm sdw} \overline\Phi \sum_y \int \!
  dx\, (-1)^y \left( \Phi_y^{\vphantom\dagger} + \Phi_y^\dagger\right),
\end{eqnarray}
where we took $\overline\Phi$ real.  

The chain mean field theory (CMFT) has now reduced the system to a
problem of decoupled chains.  It can be brought into a simple standard
form by expressing it in terms of the bosonized fields, and making the shift 
$\varphi_y \to \varphi_y + \beta y/2$, which gives finally $H_0 +
H'_{\rm MF} = \sum_y H_{\rm sG}[\theta_y,\varphi_y]$, where 
\begin{equation}
H_{\rm sG} =  \int dx ~\frac{v}{2} [(\partial_x \varphi)^2 + (\partial_x \theta)^2] - 2 \mu \cos[\frac{2\pi}{\beta} \varphi].
\label{eq:sG2}
\end{equation}
Here $\mu = \gamma_{\rm sdw} \overline\Phi$ and the self-consistency
requirement in Eq.~\eqref{eq:14} becomes
\begin{equation}
  \label{eq:17}
  \overline\Phi = \left\langle e^{i\frac{2\pi}{\beta}
      \varphi}\right\rangle_{\rm sG} =
  \left\langle \cos \frac{2\pi}\beta\varphi\right\rangle_{\rm sG}.
\end{equation}
Our notation here closely follows Refs.[\onlinecite{essler04,gnt-book}],
which describe many technical details important for the subsequent
analysis.

\subsubsection{Mass spectrum of the sine-Gordon model}
\label{sec:sine-gordon}

The excitations of the sine-Gordon model in the massive phase
($\beta^2 > \pi/2$) come in two varieties: {\em solitons and
  antisolitons}, which are domain walls connecting degenerate vacua
(minima of the cosine), and {\em breathers}, which are bound states of
solitons and antisolitons.  The number of breathers is determined by
the dimensionless parameter $\xi = 1/(8 \pi R^2 -1)$, such that $n \leq [1/\xi]$ ($[x]$ denotes closest to
$x$ integer such that $[x] \leq x$).   The minimum energy of each
breather -- the mass in the relativistic sense -- is given by the formula
\begin{equation}
\label{eq:m_n}
m_n = 2 m_s \sin[\frac{\pi}{2} \xi n] ~\text{for}~ n = 1, 2, ...[\frac{1}{\xi}],
\end{equation}
expressed here in terms of the fundamental soliton mass $m_s$. 

In the case of the spatially anisotropic triangular lattice, $\xi$ ranges
from $1/3$ at $M=0$ to $1$ at the saturation, $M=1/2$.  The breather
masses are plotted in Fig.~\ref{fig:masses} versus $M$.  For $0 < M <
0.125$, there are two breather modes.  When the magnetization is
increased to this value, the upper breather reaches the energy
of the two-soliton continuum and merges with it.  Hence, when
$0.125<M<0.5$, there is only a single breather. 

The soliton mass $m_s$ is determined by the coupling constant $\mu$ via the exact relation, \cite{zamol95}
\begin{eqnarray}
\label{eq:mu-m0}
\mu &=& \frac{v \Gamma(\frac{1}{8\pi R^2})}{\pi \Gamma(1-\frac{1}{8\pi R^2})} \Big(\frac{m_s}{v} \frac{\sqrt{\pi} \Gamma(\frac{1+\xi}{2})}
{2 \Gamma(\frac{\xi}{2})}\Big)^{2-1/(4\pi R^2)} \nonumber\\
&&\sim v (m_s/v)^{2 - 1/(4\pi R^2)} .
\end{eqnarray}
The scaling shown in the second line can be understood by simple
renormalization group arguments.  The relevant cosine operator in
\eqref{eq:sG2} grows under the RG according to $\mu(\ell) = \mu(0)
\exp[(2 - 1/(4\pi R^2))\ell]$, where $\mu(0) \equiv \mu$ is the
initial value of the coupling constant and $\ell$ is the logarithmic
RG variable, so that the running energy scale is $\epsilon \sim v
e^{-\ell}$.  The coefficient $\mu(\ell)$ reaches strong coupling at
$\ell_0$ such that $\mu(\ell_0) = v$.  Solving this for $\ell_0$, one
obtains the energy $m_s \sim v e^{-\ell_0}$, which indeed matches the
last line of \eqref{eq:mu-m0}.   The value of the the exact solution
in the {\em first} line of \eqref{eq:mu-m0}  is that it also provides with exact
numerical prefactor.

\begin{figure}[h]
\begin{center}
\includegraphics[scale=0.7]{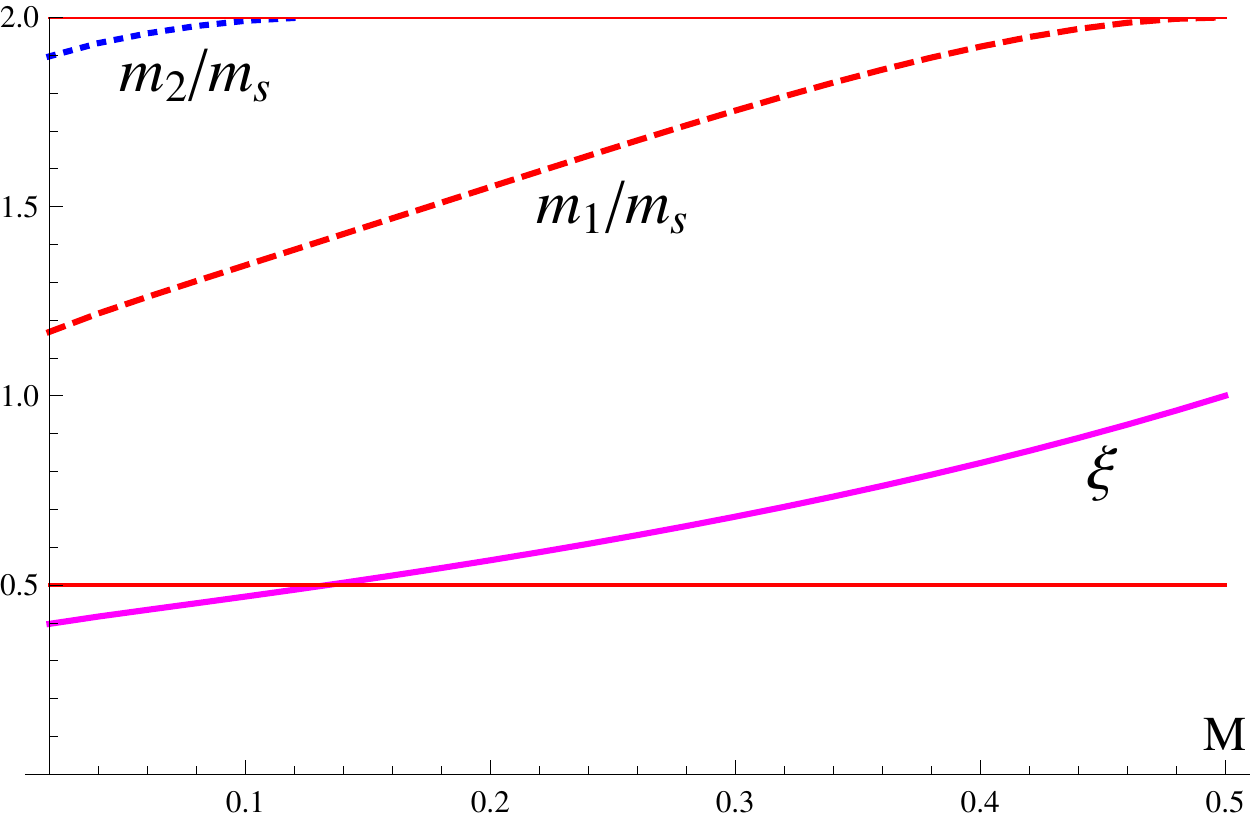}
\caption{\label{fig:masses} Plot of $\xi$ (solid (magenta) line) and breather masses $m_{1}/m_s$ (dashed (red) line) and 
$m_{2}/m_s$ (dotted (blue) line) 
as a function of magnetization $M$.
Horizontal $y=0.5$ line is used to highlight ``high magnetization'' region 
with $[1/\xi] = 1$: note that the second breather is absent there.}
\label{fig:masses}
\end{center}
\end{figure}

\subsubsection{Self-consistency}
\label{sec:self-consistency}

To determine the overall scale of the excitation spectrum, we require
the soliton mass $m_s$ or $\mu$.  This is obtained from
the self-consistency condition $\mu = \gamma_{\rm sdw} \overline\Phi$.
The expectation value defining $\overline\Phi$ is readily obtained from
the relation  
\begin{eqnarray}
\overline\Phi &=& = -\frac{1}{2} \frac{\partial F(\mu)}{\partial \mu},
\label{eq:10}
\end{eqnarray}
where $F(\mu)$ is the ground state energy density of $H_{sG}$.  Eq.~\eqref{eq:10}
follows from first order perturbation theory in changes of $\mu$.  

At the scaling level, as it is an energy {\em density}, we expect $F
\sim v m_s^2$, and using Eq.~\eqref{eq:mu-m0} one obtains
\begin{equation}
  m_s \sim v (\gamma_{\rm sdw}/v)^{2\pi R^2/(4\pi R^2 -1)}.
  \label{eq:ms-scaling}
\end{equation}
This power can be understood from RG arguments, which indicate it is
correct beyond CMFT.  Under the RG, the SDW coupling grows
according to $\gamma_{\rm sdw}(\ell)\sim \gamma_{\rm sdw} e^{(2 -
  D_{\rm sdw})\ell}$, with $D_{\rm sdw}=2\Delta_z = 1/(2\pi R^2)$,
which defines a scale $\ell_0$ by the condition that $\gamma_{\rm sdw}(\ell)$ reaches strong coupling, {\em i.e.}
becomes of order $v$.   Then using $m_s \sim v e^{-\ell_0}$, we obtain
Eq.~\eqref{eq:ms-scaling}.  

To go beyond scaling and obtain the prefactor and hence an absolute
number for $m_s$, we turn to the exact solution of the sine-Gordon
model.  The standard result in the literature is $F_{\rm standard} = -
m_s^2 \tan[\pi \xi/2]/4$.    It is, however, insufficient in the present
case due to the obvious (and unphysical) divergence of $F_{\rm
  standard}$ in the $\xi \to 1$ ($4\pi R^2 \to 1$) limit, {\em i.e.}
in the limit of $M\to 1/2$.   

This divergence is analyzed and cured,
with the help of nominally less relevant terms, in the
Appendix~\ref{ap:sG}.  We present the result here. 
To obtain the soliton mass, one first solves for $\mu$ from the equation 
\begin{eqnarray}
\label{eq:getmu}
&&(\frac{\mu}{v})^{1-\xi} = \frac{1+\xi}{8} \tan[\frac{\pi \xi}{2}] A_1^2 A_2^{1+\xi} (\frac{\gamma_{\rm sdw}}{v}) \times\\
&&\times\Big(1 - \frac{1}{8}\tan[\frac{\pi}{1+\xi}] A_1^\frac{4}{(1+\xi)} A_2^2 ~Q^\frac{(1-\xi)}{(1+\xi)} (\frac{\gamma_{\rm sdw}}{v})\Big)^{-1}. \nonumber
\end{eqnarray}
The soliton mass is then obtained as
\begin{equation}
m_s = v A_1 \Big(\frac{\mu}{v} A_2\Big)^{(1+\xi)/2}.
\label{eq:getms}
\end{equation}
This procedure
allows us to explicitly determine the soliton mass $m_s$ as a function
of $M$ for a given coupling constant $\gamma_{\rm sdw}$ of the
original spin problem.  For illustrative purposes, we plot the result
for the case of the spatially anisotropic
triangular lattice, for which $\gamma_{\rm sdw} = J' A_1^2 \sin[\pi M]$, with $J'/J = 0.5$ (chosen arbitrarily) in
Figure~\ref{fig:delta}. 

\begin{figure}[h] \begin{center} \includegraphics[scale=0.65]{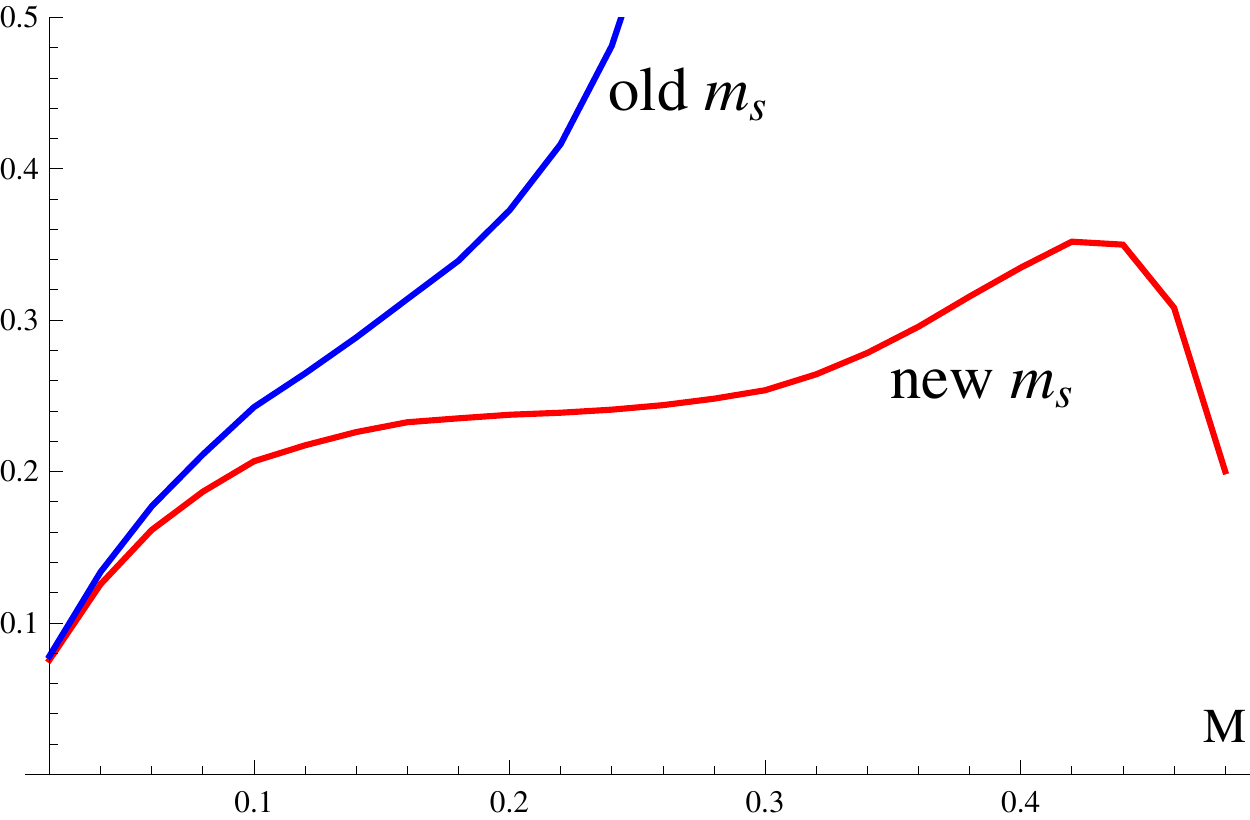} \caption{\label{fig:delta}
      Plot of soliton mass $m_s/v$ as function of magnetization $M$
      for $J'=0.5 J$.  ``Old $m_s$" (blue curve) is obtained using
      $F_{\rm standard}$, without correcting for $\xi\to 1$
      divergence.  ``New $m_s$" (red curve) is corrected result
      \eqref{eq:getms}, which is obtained using $F_{\rm new}$ from the
      Appendix~\ref{ap:sG}.  For the spatially anisotropic triangular
      lattice the SDW phase, for which $m_s$ is calculated here, is
      the ground state of the 2d problem in the interval $0 < M
      \lesssim 0.3$.  At higher $M$ the SDW is replaced by the cone
      phase.  Note that, in the limit $J'/J \rightarrow 0$, at fixed
      $M$, the two curves converge to one another (in fact the ratio
      of $m_s$ calculated in both fashions converges to one).
    }  \label{fig:delta} \end{center} \end{figure}

\subsection{Spin susceptibilities}

The aim of this subsection is to show how the excitations described in
the prior section, which are excitations already on a single chain,
and the collective modes, which appear only when the full 2d dynamics
are considered, appear in the physical dynamical susceptibilities,
i.e. the components of the dynamical structure factor measured in
inelastic neutron scattering.  Formally these are defined as the
linear response quantities,
\begin{equation}
  \label{eq:18}
  X_{\mu\nu}({\bf k},\omega) = \left. \frac{\delta S^\mu({\bf
        k},\omega)}{\delta h^\nu({\bf k},\omega)}\right|_{{\bf h}({\bf k},\omega)=0},
\end{equation}
where ${\bf h}$ is an oscillating infinitesimal applied Zeeman field
at wavevector ${\bf k}$ and frequency $\omega$.  By the usual linear
response theory, this is {\em minus} the retarded correlation function
of spin operators
\begin{equation}
  \label{eq:19}
  X_{\mu\nu}({\bf k},\omega) \sim i\int_0^\infty \! dt\, e^{(i\omega-\epsilon)t}
  \left\langle \left[ S^\mu({\bf k},t),
    S^{\nu}(-{\bf k},0)\right]\right\rangle ,
\end{equation}
where $\epsilon=0^+$.  

We distinguish two types of susceptibilities.  The {\em longitudinal}
susceptibility describes the dynamical correlations of spin components
$S^z$ along the applied field and the SDW polarization.  Using the
bosonization rule of Eqs.~\eqref{eq:11} and \eqref{eq:1}, we see that this is related to
correlations of the SDW order parameter $\Phi$.  Hence we define the
bosonized equivalent, $\chi^{\rm zz}$, of the longitudinal susceptibility
\begin{equation}
  \label{eq:20}
  X_{zz}({\bf
  k}=(k_{\rm sdw}+q,\pi + q_y),\omega)  \sim \chi^{\rm zz}(q,q_y,\omega) ,
\end{equation}
and hence
\begin{eqnarray}
  \label{eq:31}
\chi^{\rm zz}(q,q_y,\omega) & = & i\int_0^\infty \!
  \!\! \!\! dt\!\! \int\!\! dx \sum_y \,e^{i q x+iq_y y+(i\omega-\epsilon)    t}\nonumber \\
  && \times \left\langle \left[\Phi_y(x,t),
    \Phi_0^\dagger(0,0)\right]\right\rangle .
\end{eqnarray}
Note that in $\chi^{zz}(q,q_y,\omega)$, $q$ gives the {\em shift} of
the momentum along the chain from the SDW one, i.e. $k_x=k_{\rm sdw} + q$, 
while $q_y$ is measured from $\pi$ due to the shift of $\varphi$ field by $\beta y/2$ made in deriving \eqref{eq:sG2}. 
Moreover, the continuum formula in Eq.~\eqref{eq:31} describes only
the contributions to the susceptibility at low energy near $k_x=k_{\rm sdw}, k_y = \pi$.
Other contributions may apply elsewhere.  For example, 
contribution from the vicinity of $k_x= - k_{\rm sdw}, k_y = \pi$ is described by the hermitian conjugate of the expression in Eq.~\eqref{eq:31}, while in that
near $k_x=0$, the operator $\partial_x\varphi_y$ in Eq.~\eqref{eq:2} or
Eq.\eqref{eq:11} contributes.  We neglect it here because, since this operator has larger scaling
dimension than $\Phi_y$, it gives a subdominant contribution in
the sense of smaller integrated weight in $X_{zz}$ (i.e. the weight
near $k_x=0$ is smaller than that near $k_x=\pm k_{\rm sdw}$).  

The {\em transverse} susceptibility describes the spin components
$S^x$,$S^y$ normal to the field and the SDW axis.  Using the
bosonization rule in Eqs.~(\ref{eq:2},\ref{eq:10a}), we find
\begin{eqnarray}
  \label{eq:21}
  X_{xx}({\bf
  k}=(\pi+q,k_y),\omega)  & = &  X_{yy}({\bf
  k}=(\pi+q,k_y),\omega)  \nonumber \\
& \sim & \chi^{\rm xy}(q,k_y,\omega) ,
\end{eqnarray}
with 
\begin{eqnarray}
\label{eq:22}
\chi^{\rm xy}(q,q_y,\omega) & = & i\int_0^\infty \!
  \!\! \!\! dt\!\! \int\!\! dx \sum_y \,e^{i q x+i k_y y+(i\omega-\epsilon)    t}\nonumber \\
  && \times \left\langle \left[\Psi_y(x,t),
    \Psi_0^\dagger(0,0)\right]\right\rangle .
\end{eqnarray}
As for the longitudinal one, we have defined the continuum transverse
susceptibility $\chi^{\rm xy}(q,q_y,\omega)$ in such a way that $q$
gives a shift in momentum relative to some offset, but with a {\em
  different} offset from the one used in the longitudinal
susceptibility.  Here $q_x = \pi + q$, i.e. $k_{\rm sdw} \rightarrow \pi$ on
passing from the longitudinal to transverse susceptibility.  This
difference originates from the distinct momenta of singular response
of a one dimensional spin system in the two channels.  It must be
noted that while we can study this object, defined
by Eq.~\eqref{eq:22}, also for the case of the SDW formed from SN
chains, in that case it is {\em not} the true transverse spin
susceptibility.  Due to the definition of $\Psi$ for the SN case, it
instead represents the nematic susceptibility.  

In the following, we obtain these quantities using the RPA approximation, which
expresses these 2d dynamical susceptibilities in terms of the 1d
dynamic susceptibilities of the individual decoupled chains we
obtained in the CMFT approximation.  

\subsubsection{Susceptibilities of the sine-Gordon model}

We now obtain the 1d dynamical susceptibilities.  These are by
construction independent of $q_y$.  According to
bosonization, the longitudinal and transverse susceptibilities are
related to correlations of exponentials of $\varphi$ and $\theta$
fields, respectively.  The corresponding correlations of the
sine-Gordon model may be calculated via the form-factor expansion
which is described in great detail in Ref.~\onlinecite{essler04}.
Here we present key results from this reference as adapted for our
needs.

{\em Longitudinal susceptibility:} The longitudinal susceptibility is obtained from the two-point
correlation function of $\Phi$ in Eq.~\eqref{eq:31}. 
What excitations are created by this effective longitudinal spin
operator? Since $\Phi$ is local in $\varphi$ (see Eq.~\eqref{eq:10a}),
it cannot generate topological excitations with non-zero soliton
number.  Instead, acting on the ground state, it generates gapped
excitations corresponding to breathers, unbound soliton-antisoliton
pairs, and also higher energy states such as multiple breather states.
The largest contribution, however, comes simply from the first
breather $B_1$ (in the notation of Ref.~\onlinecite{essler04}).  In
the approximation in which only this excitation contributes, the
longitudinal susceptibility has a single simple pole,
\begin{equation}
\chi_{1d}^{\rm zz}(q,\omega) = \frac{C_z Z_z}{m_1^2 + v^2 q^2 -\omega^2 - i \epsilon}.
\label{eq:chi1zz}
\end{equation}
The mass $m_1$ is given by $n=1$ in \eqref{eq:m_n}. Note that in the
whole magnetization range $0 < M < 1/2$, when $1/3 < \xi < 1$, the
first breather's mass exceeds that of the soliton, $m_1 > m_s$.  The
residue $Z_z = v (m_s/v)^{1/(2\pi R^2)}$ is determined by the soliton
mass $m_s$, while the factor $C_z$ collects all numerical coefficients and
depends smoothly on the magnetization $M$. The second breather $B_2$
does not contribute because it only connects states of the same parity while
$S^z(0,0)$ is odd under parity.

The continuum soliton-antisoliton states become available for $\omega
\geq 2m_s$.  In the form factor expansion of
Ref. \onlinecite{essler04}, this contribution was denoted $F_{+-}$.
We consider energies close to the threshold, $s-2m_s \ll 2m_s$, where
$s=\sqrt{\omega^2 - v ^2 q^2}$.  With some analysis of formula in that
reference, we find that the contribution $F_{+-}$ to the dynamic
structure factor of the single chain starts smoothly as $\sqrt{s-2
  m_s} ~\Theta(s-2 m_s)$.  This is in accord with the general behavior
expected for the two particle contribution to correlation functions of
one dimensional systems in the situation where the particles
experience attractive interactions (which must be the case here since
bound states (breathers) form).  In general, for $s>2m_s$, all other
contributions will occur inside the two soliton continuum, and we
expect that mixing with the continuum will remove any sharp features
at higher energies (though this mixing may be controlled by deviations
from integrability).  The end result is that Eq.~\eqref{eq:chi1zz}
should be supplemented by the continuum contribution for $s>2m_s$,
which extends smoothly to higher energies.

{\em Transverse susceptibility:} The transverse susceptibility is
obtained from correlations of $\Psi$ as in Eq.~\eqref{eq:22}.  The field
$\Psi_y =e^{-i\beta\theta_y}$ is {\em not}
local in the $\varphi$ variables, and indeed $\theta$ can be expressed as an
integral of the canonical momentum conjugate to $\varphi$.
Consequently, it creates soliton and antisoliton defects in
$\varphi$, and some algebra shows that it changes the topological charge $Q_{\rm charge}
= \beta^{-1} \int dx ~\partial_x \varphi$ by $\pm 1$.   Hence the
lowest energy contribution to the transverse susceptibility is simply
that of single solitons, and again has a pole form.  Thus 
\begin{equation}
\chi_{1d}^{\rm xy}(q,\omega) = \frac{C_{\rm xy} Z_{\rm xy}}{m_s^2 + v^2 q^2 -\omega^2 - i \epsilon}.
\label{eq:chi1perp}
\end{equation}
Here $Z_{\rm xy} = v (m_s/v)^{2\pi R^2}$ while $C_{\rm xy}$ includes
all numerical coefficients and smooth dependence on $\beta=2\pi R$ and
magnetization $M$.  Note that $m_s<m_1$ so that the first onset of
spectral weight in the chain occurs here in the transverse
correlation function rather than the longitudinal one.   

Corrections to this form account for multi-particle contributions to
$\chi^{\rm xy}$.   These can be of soliton-breather ($s-$B$_1$) and of
soliton-soliton-antisoliton ($s-s-\bar{s}$) types, as is schematically
shown in eq.3.73 of Ref.\onlinecite{essler04}.  They appear at energy
$m_s+ m_1 > 2 m_s$ and $3 m_s$.  Thus the continuum contribution
for the transverse susceptibility occurs above the one for the
longitudinal one.  We do not pursue it further here.  The spectral content of equations \eqref{eq:chi1zz} and
\eqref{eq:chi1perp} is schematically depicted in
Fig.~\ref{fig:SG}.

It is instructive to compare the excitation structure found here with
the ``dual'' sine-Gordon problem which has been frequently discussed
in other problems of one-dimensional magnetism, in which the ordering
is transverse, so $\cos[2\pi \varphi/\beta]$ in \eqref{eq:sG2} is
replaced by $\cos[\beta\theta]$.  In that case,\cite{essler03} the
parameter $\xi$ ranges from $1/3$ at zero magnetization, $M=0$, to
$1/7$ at $M=1/2$, resulting in many more breathers (up to 7) peeling
off of the soliton-antisoliton continuum with an {\em increasing}
number with increasing magnetization. In parallel with this, the
spectral composition of different excitations branches changes
accordingly: the breathers contribute near momentum $\pi$, while
solitons (antisolitons) contribute near momentum $\pi (1 + 2 M)$ ($\pi
(1 - 2 M)$) -- see for example Fig.1 of Ref.\onlinecite{essler03}.

\subsubsection{Susceptibility of 2d SDW phase}
\label{sec:rpa}

The single chain approximation is not sufficient for describing
two-dimensional (2d) spin correlations.  At the single chain level,
all spin excitations have a gap, there is no dispersion transverse to
the chains (i.e. dependence upon $q_y$), and there are no Goldstone
(spin wave) modes.  These deficiencies are easily fixed, however, with
the help of a simple random-phase approximation (RPA) in the
interchain couplings, as suggested by Schulz and developed in great
details by Essler and Tsvelik.

We apply the RPA approximation directly to the continuum problem of
correlations of $\Phi$ and $\Psi$.  This gives expressions for the 2d
susceptibilities directly from the single-chain susceptibilities,
$\chi_{1d}^{{\rm zz},{\rm xy}}$ described above:
\begin{equation}
  \chi_{\rm 2d}^\alpha(q,k_y,\omega) = \frac{\chi_{1d}^\alpha(q,\omega)}{1 + 2
    \gamma_\alpha(q,k_y) \chi_{1d}^\alpha(q,\omega)}. 
\label{eq:chi-rpa}
\end{equation}
Here $\alpha = {\rm zz}, {\rm xy}$ describes the two channels,
$\gamma_\alpha(q,k_y)$ is the Fourier transform of the interchain
interaction in the $\alpha$-channel:
\begin{eqnarray}
  \label{eq:23}
  \gamma_{\rm zz}(q,k_y) & = &  \gamma_{\rm sdw} \cos k_y, \\
  \gamma_{\rm xy}(q,k_y) & = & \left[\gamma_{\rm xy} - q \gamma'_{\rm
      xy}\right]\cos k_y.
\end{eqnarray}
The parameters $\gamma_{\rm sdw}$, $\gamma_{\rm xy}$, and
$\gamma'_{\rm xy}$ are collected for convenience in Table~I.  Using
them, and
Eqs.~(\ref{eq:23},\ref{eq:chi-rpa},\ref{eq:chi1perp},\ref{eq:chi1zz}),
one can obtain the two dimensional susceptibility for any of the three
models discussed here. 

As an example, we discuss this now in some detail for the case of the
spatially anisotropic triangular antiferromagnet.  Applying
Eq.~\eqref{eq:23} and Table~I, we obtain $\gamma_{\rm zz}(q,k_y) =  J' 
A_1^2 \sin(\pi M) \cos k_y$ and $\gamma_{\rm xy}(q,k_y) = -
\frac{1}{2}J' q A_3^2 \cos k_y$.  We see that $\gamma_{\rm xy} \ll
\gamma_{\rm zz}$ owing to the additional factor of $q \ll 1$ in this
term, which ultimately arose from inter-chain frustration.  

Hence in the ordered two-dimensional SDW state
\begin{eqnarray}
\label{eq:2d-zz}
&&\chi_{\rm 2d}^{zz}(q,k_y,\omega)=
\Big((\chi_{1d}^{zz}(q,\omega))^{-1} + 2 \gamma_{\rm zz}(k_y)\Big)^{-1}  \\
&&= \frac{C_z Z_z}{\Big[ (m_1^2 + 2C_z Z_z J' A_1^2 \sin(\pi M)\cos[k_y]) + v^2 q^2 -\omega^2\Big]}.
 \nonumber
\end{eqnarray}
As written, this expression is characterized by a finite, albeit
renormalized and $k_y$-dependent, gap in the spin excitation spectrum,
$m_{\rm sdw}^2 = m_1^2 + 2C_z Z_z J' A_1^2 \sin(\pi M)\cos[k_y] \neq 0$ and
does not seem to describe a gapless phason mode. 
This shortcoming is of
course due to the approximate nature of the RPA expression
\eqref{eq:chi-rpa}. Since the phason is a Goldstone mode which is
required by the very existence of the 2d SDW order, we follow Schulz
and simply require that the gap must close at some appropriate $k_y$.
Clearly for $C_z > 0$ this happens at $k_y = \pi$.  This reflects the
preference of SDWs on adjacent chains to order out of phase due to
repulsive (antiferromagnetic) interactions between them.

To check the consistency of this procedure we need to make sure that
both terms in the expression for $m_{\rm sdw}^2$ scale in the same way
with $J'/J$ -- and this is exactly what we find. While $m_1^2 \sim
(J')^{4\pi R^2/(4\pi R^2 -1)}$ in accordance with
\eqref{eq:ms-scaling}, it is also easy to see that the interchain term
$J' Z_z \sim (J')^{1 + 1/(4\pi R^2-1)}$ follows the same
power law. Thus the two terms are of the same order and our {\em
  requirement} $m_1^2 = 2C_z Z_z J' A_1^2$ simply fixes the
overall numerical coefficient $C_z$ of the longitudinal
susceptibility.

Hence, in the vicinity of ordering momentum ${\bf k}=(k_{\rm sdw},
\pi)$, we have, with $k_x = k_{\rm sdw}+q$ and $k_y = \pi + q_y$,
\begin{equation}
  \chi_{\rm 2d}^{zz}(q,\pi+q_y,\omega) \sim \frac{Z_{\rm zz;2d}}{(v^2 q^2 + v_\perp^2 q_y^2) - \omega^2} ,
\label{eq:2d-phason}
\end{equation}
with $Z_{\rm zz;2d} = m_1^2/(2 J' A_1^2)$, when $q, q_y \ll 1$.
The phason has linear dispersion
\begin{equation}
\omega = \sqrt{v^2 q^2 + v_\perp^2 q_y^2}
\label{eq:disp-phason}
\end{equation}
with strongly anisotropic velocity. Its
transverse (inter-chain) velocity $v_\perp = \sqrt{m_1^2/2} \sim J (J'/J)^{2\pi R^2/(4\pi R^2 -1)}$ is much
smaller than $v \sim J$.  

In the transverse $({\rm xy})$ channel we have instead
\begin{eqnarray}
&&\chi_{\rm 2d}^{\rm xy}(q,k_y,\omega) = \nonumber\\
&&\frac{C_{\rm xy} Z_{\rm xy}}{\tilde{m}_s^2(k_y) + \big(v q - C_{\rm xy} Z_{\rm xy} J' A_3^2 \cos[k_y]/2v^2\big)^2 - \omega^2} .
\label{eq:chi-perp}
\end{eqnarray}
Here 
\begin{equation}
\label{eq:41}
\tilde{m}_s^2(k_y) = m_s^2 - (C_{\rm xy} Z_{\rm xy} J' A_3^2\cos[k_y]/2v)^2
\end{equation}
is the renormalized gap which depends on the transverse momentum $k_y$.

Note that the second term in the renormalized gap is
negative, so there is the potential for an instability in this
expression, if the negative correction becomes larger than the
positive $m_s^2$ term.  Let us examine the relative magnitude of the
two terms.  Unlike those considered above for the longitudinal
susceptibility, here they scale differently with $J'/J$.  The $k_y$-dependent
correction, $(Z_{\rm xy} J'/v)^2$, scales as $(J'/J)^{\alpha_2}$ with
$\alpha_2 = 2 + 2 (2\pi R^2)^2/(4\pi R^2 -1)$, while $m_s^2$ scales as
$(J')^{\alpha_1}$ with $\alpha_1 = 4\pi R^2/(4\pi R^2 -1)$.
Importantly $\alpha_1 < \alpha_2$ at low magnetization where $2\pi R^2
\approx 1$.   Hence when $\alpha_1<\alpha_2$, $\tilde{m}_s(k_y)$ is parametrically dominated by
the first term and is positive for all $k_y$.   As the compactification
radius diminishes with increasing magnetization, the exponent
$\alpha_2$ decreases as well and at some critical point becomes equal
to $\alpha_1$. This happens when $2 \pi R^2 = (\sqrt{5}-1)/2$, which
takes place at approximately $M =0.3$.   This signals an instability
of the SDW phase.  Recall in Sec.~\ref{sec:spat-anis-triang} we
derived a condition on the formation of the SDW phase, $D_{\rm sdw}=2\Delta_z
< D_{\rm xy}=1+2\Delta_\perp$.  Straightforward algebra shows the two
conditions to be indentical, thus strikingly showing the consistency
of the CMFT+RPA theory with general RG arguments!   

It is clear that at this critical point the gap closes, at 
$k_y = 0,\pi$, and the system enters magnetically ordered cone state where
spin components transverse to the external field acquire a finite
expectation value. However below such a magnetization the SDW phase is
stable and transverse spin fluctuations are massive (but coherent,
i.e. single-particle like), as \eqref{eq:chi-perp} shows.  The minimal
gap occurs at momenta $\pm q' = \pm C_{\rm xy} Z_{\rm xy} A_3^2 J'/2v^2$,
which describes a small shift away from the commensurate point.  
In terms of the full 2d momentum, the minima are at ${\bf k}_1 = (\pi - q', \pi)$ and ${\bf k}_2 = (-\pi + q', 0)$.

\subsubsection{Response near $k_x =0$}

To describe $k_x \approx 0$ region of the Brillouin zone, we need to account for the so far
neglected less relevant terms of the mode expansion in Eq.\eqref{eq:2} and Eq.\eqref{eq:1}.
For $S_y^z(x)$ this is given by the derivative term $\beta^{-1} \partial_x \varphi_y(x)$ in Eq.\eqref{eq:2},
while $S_y^+(x)$ has additional contributions at momenta $\pm 2\pi M$ which read
\begin{eqnarray}
S_y^+(x) &=& \frac{i A_2}{2} e^{- i 2\pi M x} e^{i\beta \theta} e^{i\frac{2\pi}{\beta} \varphi} + \text{h.c.}
\label{eq:100}
\end{eqnarray}
Observe that $S_y^+(x)$ can be written, with the help of \eqref{eq:10a}, as
\begin{equation}
S_y^+(x) = \frac{- i A_2}{2} e^{i 2\pi M x} \Phi_y(x) e^{i\beta \theta} +  \text{h.c.}
\label{eq:101}
\end{equation}
This form makes it clear that the main effect of the SDW ordering, as described by the chain mean-field approximation
\eqref{eq:15} and \eqref{eq:14}, is captured by the replacement 
$\Phi_y(x) \to \langle \Phi_y(x) \rangle = \bar{\Phi} (-1)^y $. Hence 
\begin{equation}
S_y^+(x) \to \frac{- i A_2}{2} e^{i 2\pi M x + i \pi y} \bar{\Phi} e^{i\beta \theta} +  \text{h.c.},
\label{eq:102}
\end{equation}
which makes it proportional to $\psi_y(x)$ in \eqref{eq:2} -- but located near $k_x = \pm 2\pi M$
instead of $\pi$.

Thus transverse spin susceptibility in the vicinity of momenta ${\bf k} = (\pm 2\pi M, \pi)$ is given by
Eq.\eqref{eq:chi-perp} with $k_x = \pm 2\pi M + q$ and $k_y = \pi + q_y$ and 
with the renormalized residue $Z_{\rm xy} \to Z_{\rm xy}  \bar{\Phi}^2$. Observing that SDW order parameter $\bar{\Phi} \ll 1$
we conclude that the total spectral weight of this contribution is much smaller than that from 
the momentum $k_{\rm sdw}$, Eq.\eqref{eq:chi-perp}. Notice that near
saturation, the momentum $2\pi M$ is closer to $\pi$ than to $0$, and certainly can be
to the right of the SDW wavevector $\pi(1-2M)$.  

Consideration of the longitudinal susceptibility near $k_x =0$ requires more care.
Mean-field Hamiltonian \eqref{eq:sG2} implies that 
\begin{equation}
\chi_{1d}^{\rm zz}(k_x\approx 0,\omega) \sim \frac{\tilde{C}_z k_x^2}{m_1^2 + v^2 k_x^2  -\omega^2}.
\label{eq:103}
\end{equation}
(Similarly to Eq.\eqref{eq:chi1zz} the second breather, of mass $m_2$, does not contribute here to do oddness of 
$\partial_x \varphi_y(x)$ under parity transformation.) Observing that inter-chain coupling of the uniform components
$\partial_x \varphi_y$ of $S_y^z$ is given by $2 J' \cos[k_y]$ (it is not frustrated), 
RPA approximation \eqref{eq:chi-rpa} would then suggest 
that two-dimensional susceptibility has the form
\begin{eqnarray}
&&\chi_{2d}^{\rm zz, {\bf RPA}}(k_x\approx 0,k_y,\omega) \sim \nonumber\\
&&\frac{\tilde{C}_z k_x^2}{m_1^2 + v^2 k_x^2(1 + a J' \cos[k_y]/v)  -\omega^2} ~[\text{{\bf wrong!}}],
\label{eq:104}
\end{eqnarray}
where $a$ is numerical coefficient. Thus RPA predicts gapped excitation with $\omega \sim m_1$ 
which is not correct. The basic reason for this
is that RPA ``does not know" about the gapless phason mode \eqref{eq:disp-phason} - recall that in going from \eqref{eq:2d-zz}
to \eqref{eq:2d-phason} we have imposed the gaplessness condition by hand.

On the other hand, the Ginzburg-Landau action of Appendix~\ref{ap:phason}
does capture this crucial property of the SDW ground state properly:
Eq.\eqref{eq:ap11} shows that $\partial_x \varphi_y(x) = \partial_x \Phi(x,y)$ which,
in view of the phason action Eq.\eqref{eq:ap12}, leads to the desired result,
\begin{equation}
\chi_{2d}^{\rm zz}(k_x,k_y,\omega) = 
\frac{v k_x^2/\beta^2}{v^2 k_x^2 + v_\perp^2 k_y^2  -\omega^2} ,
\label{eq:105}
\end{equation}
where the transverse phason velocity $v_\perp/v \sim  (\gamma_{\rm sdw}/v)^{2\pi R^2/(4\pi R^2 - 1)} \ll 1$,
according to \eqref{eq:disp-phason} and \eqref{ap:v-perp}, and
$k_x,k_y \ll 1$.  Taking the
imaginary part (using the usual $i0^+$ prescription), we find
\begin{equation}
  \label{eq:27}
  {\rm Im}\, \chi_{2d}^{\rm zz}(k_x,k_y,\omega) \sim
  \frac{vk_x^2}{\sqrt{v^2 k_x^2 + v_\perp^2 k_y^2}}\delta(\omega - \sqrt{v^2 k_x^2 + v_\perp^2 k_y^2}).
\end{equation}
Eq.\eqref{eq:27} demonstrates that acoustic 2d phason mode can be
observed near ${\bf k} \approx 0$, in addition to the vicinity of $\pm
\pi (1 - 2M)$ (Eq.\eqref{eq:2d-phason}).  It has weight that vanishes
linearly as $k \rightarrow 0$ but is also anisotropic: it vanishes on
the line ${\bf k} = (0,k_y)$.

{\em We now summarize the results for the spatially anisotropic triangular
lattice}.  The above discussion shows that the onset of spectral weight
in the two-dimensional susceptibility $\chi_{\rm 2d}$ occurs as
well-defined collective modes, in both the longitudinal and transverse
channels.  They are descended from the breather and soliton
excitations of the sine-Gordon model, respectively.  The outlined approach
predicts not only the dispersion of these modes, but also their
spectral weight.  Though we did not discuss this in any detail,
the RPA also allows an analysis of the continuum spectrum which
appears in (and dominates) the higher energy region.

Further analysis, summarized in Appendix~\ref{ap:plateau}, is required
to describe {\em commensurate} SDW order which becomes pinned to the lattice 
by weak multi-particle umklapp processes. In this case, which corresponds
to a two-dimensional magnetization plateau state, the phason mode acquires a gap
in the spectrum.  See Eq.~\eqref{eq:2d-phason-plateau} and surrounding
discussion for details.

\section{Spin Nematic}
\label{sec:nematic}

The aim of this section is mainly to repeat the considerations of the previous one for
the case of a spin nematic (SN), discussing the features of the
corresponding excitation spectrum.  However, we first present a
``derivation'' via bosonization of the effective quasi-1d theory for
a spin nematic, relevant to experiment.

\subsection{1d nematic}
\label{sec:1d-nematic}

A case for the spin nematic state has been made in the material
LiVCuO$_4$.  It consists of weakly coupled spin chains with
significant nearest and second-nearest neighbor Heisenberg exchange,
i.e. $J_1 - J_2$ chains.  Here the nearest-neighbor interaction is
ferromagnetic $J_1 < 0$, and the second neighbor $J_2>0$ is
antiferromagnetic, and we take $J_2 \gg |J_1|$.   In this limit, one
may naturally view each chain as a ``zig-zag ladder'' of the two
sub-chains formed by even and odd sublattices (and connected by
$J_2$), cross-coupled by $J_1$, see Figure~\ref{fig:geometry}.  One may thereby bosonize the two
sub-chains separately, introducing a doubled set of bosonized fields $\varphi_{y,{\rm
    odd}}, \theta_{y,{\rm odd}}$ and $\varphi_{y,{\rm
    even}},\theta_{y,{\rm even}}$ for each chain $y$.  

The nematic state arises, in this picture, from the SDW coupling
between the two sub-chains, which can dominate due to the fact that
the zig-zag coupling frustrates the XY interactions.  The sub-chain SDW coupling
takes the bosonized form
\begin{eqnarray}
  \label{eq:5}
  H_{\rm sub-chain} & \sim & \sum_y \int \! dx\, J_1 \sin[\pi M]
  \cos[\frac{2\pi}{\beta} (\varphi_{y, {\rm odd}} - \varphi_{y, {\rm
      even}})] \nonumber \\
  & \sim & \sum_y \int \! dx\, J_1 \sin[\pi M] \cos[\sqrt{2} \varphi_y^{-}/R],
\end{eqnarray}
where
\begin{equation}
  \varphi_y^\pm = (\varphi_{y, {\rm odd}} \pm \varphi_{y, {\rm even}})/\sqrt{2} .
\end{equation}
At not too low fields, $\pi M$ is close to $\pi/2$, and this
interaction is large, pinning the relative mode $\varphi_y^{-}$
strongly.  As a result, the conjugate field $\theta_y^{-}$ is highly
fluctuating, rendering harmonics of it quantum disordered on rather
short length scales. 

These observations correspond to the formation of the 1d nematic.
This can be seen by expressing the spin operators in the $\pm$ basis:
\begin{eqnarray}
S^z_y(x) & \sim & A_1 {\rm Im} \left[ e^{i\frac{2\pi}{\sqrt{2}\beta}
  \varphi_y^+(x)} e^{ i (-1)^x \frac{2\pi}{\sqrt{2}\beta} \varphi_y^-(x)}
e^{- i k_{\rm sdw} x}\right], \nonumber \\
S^+_y(x) & \sim &  (-1)^x A_3 e^{i \frac{\beta}{\sqrt{2}}
  \theta_y^+(x)}  e^{i (-1)^x \frac{\beta}{\sqrt{2}} \theta_y^-(x)},
\end{eqnarray}
where the $(-1)^x$ factors {\sl inside} the exponentials arise from
the decomposition into even and odd sub-chains.   One sees that, due
to the presence of the $\theta_y^-(x)$ field in the exponential,
$S^+_y(x)$ is quantum disordered, and has therefore very short-range
correlations.  However, one may construct the nematic operator, 
\begin{equation}
  \label{eq:6}
  T_y^+ = S^+_y(x) S^+_y(x+1) \sim e^{i \sqrt{2}\beta \theta_y^+(x)},
\end{equation}
for which the $\theta_y^-(x)$ field cancels, and which therefore has
power-law correlations.  Note that of course $S^z_y(x)$ also has
power-law correlations, as it contains not $\theta_y^-(x)$ but the conjugate field
$\varphi_y^-(x)$, which can be set to zero at low energy.  

Connecting with the discussion in Sec.~\ref{sec:spin-nematic-chains},
we identify $T_y^+ \sim \psi_y$, and hence $\theta_y =
\sqrt{2}\theta_y^+$ and $\varphi_y = \varphi^+_y/\sqrt{2}$ (the latter
normalization preserves the commutation relations).  In these
variables, the spin operators become
\begin{eqnarray}
S^z_y(x) & \sim & A_1 {\rm Im} \left[ e^{i\frac{2\pi}{\beta}
  \varphi_y(x)} e^{ i (-1)^x \frac{2\pi}{\sqrt{2}\beta} \varphi_y^-(x)}
e^{- i k_{\rm sdw} x}\right], \nonumber \\
S^+_y(x) & \sim &  (-1)^x A_3 e^{i \frac{\beta}{2}
  \theta_y(x)}  e^{i (-1)^x \frac{\beta}{\sqrt{2}} \theta_y^-(x)}.
\label{eq:8}
\end{eqnarray}

\subsection{Competition between 2d SN and paired SDW}
\label{sec:comp-betw-2d}

A two dimensional nematic can be stabilized by coupling between
chains, but this interaction can also stabilize a paired SDW.
Consider the interchain interaction $J'$ of the transverse spin
components, which we presume acts in an unfrustrated way, coupling
even sublattice to even sublattice, and odd sublattice to odd
sublattice.  Then
\begin{eqnarray}
H_3 & = &  \sum_y \sum_{a} \int dx\, J' \cos[\beta(\theta_{y,a} -
\theta_{y+1,a})] \\
& = &  \sum_y \int dx \, 2J' \cos[\frac{\beta}{\sqrt{2}}(\theta_y^+ -
\theta_{y+1}^+)] \cos[\frac{\beta}{\sqrt{2}}(\theta_y^- - \theta_{y+1}^-)] \nonumber \\
& = &  \sum_y \int dx \, 2J' \cos[\frac{\beta}{2}(\theta_y -
\theta_{y+1})] \cos[\frac{\beta}{\sqrt{2}}(\theta_y^- - \theta_{y+1}^-)]. \nonumber
\end{eqnarray}
In the first line $a$ sums over even and odd sub-chains.  In the last
line, we have re-expressed the interaction in terms of the nematic
phase $\theta_y$ defined above and in
Sec.~\ref{sec:spin-nematic-chains}.  Due to the
presence of the fluctuating $\theta_y^-$ field, the above operator has
only short range correlations and is highly irrelevant.  It, however,
generates a nematic interaction, which can be easily
obtained by integrating out the $\theta_y^-$ field in a cumulant
expansion.  The result has the form \cite{sato}
\begin{eqnarray}
\label{eq:nem-H}
H_{\rm nem} &\sim& \sum_y \int dx \, (J'^2/J_1) \cos[\beta(\theta_y - \theta_{y+1} )] \\
& \sim& \sum_y \int dx \, (J'^2/J_1) ~[\psi_y^+(x) \psi_{y+1}^-(x) + {\rm h.c.}]. \nonumber
\end{eqnarray}
This involves only the slowly varying $\theta_y$ fields, and indeed
has the same mathematical form as the XY interaction between
non-frustrated chains, Eq.~\eqref{eq:Hp-non-fr}, with $\gamma_{\rm xy}
\sim J'^2$.  The nematicity of the problem is encoded in the
definition of $\theta_y$.  If we were to try to generate
Eq.~\eqref{eq:nem-H} directly microscopically (i.e. with a coefficient
proportional to a microscopic coupling), it would require a four-spin
interaction, e.g.
\begin{equation}
 H_{\rm nem} \sim  (J'^2/J_1) \sum_y \int dx ~[T_y^+(x) T_{y+1}^-(x) + {\rm h.c.}].
\end{equation}

As discussed in Section~\ref{sec:spin-nematic-chains}, nematic interchain interaction \eqref{eq:nem-H}
competes against the direct $S^z - S^z$ (density - density) interaction 
\begin{equation}
H_{\rm sdw} \sim \sum_y \int dx \, J' \sin[\pi M]  \cos[\frac{\sqrt{2}\pi}{\beta}(\varphi^+_y - \varphi^+_{y+1} )] ,
\label{eq:nem-sdw}
\end{equation}
which drives the system of nematic spin chains towards the
longitudinal SDW state. It is interesting to note that the
competition between `dual' magnetic orders \eqref{eq:nem-H} and
\eqref{eq:nem-sdw} is quite similar to that between superconducting
and charge-density wave orders in itinerant charge systems
\cite{Jaefari2010}.

As usual, relative importance of the two
competing interactions can be estimated by comparing their scaling
dimensions. Scaling dimension of the nematic interchain interaction
$D_{\rm nem} = 2 \cdot (\sqrt{2}\beta)^2/(4\pi) = 4\pi R^2 \in (2,1)$
ranges from 2 at zero magnetization to 1 near the saturation.  Scaling
dimension of the SDW interaction $D_{\rm sdw} = 2 \cdot
(\sqrt{2}\pi/\beta)^2/(4\pi) = 1/(4 \pi R^2) \in (1/2, 1)$.  We see
that $D_{\rm sdw} < D_{\rm nem}$ for all magnetization values, except
the very vicinity of the saturation where the two coincide within our
crude approximation which neglects less relevant and marginal
interchain interactions which do weakly modify scaling dimensions
$D_{\rm nem/sdw}$ of the leading terms. In addition, the nematic
interaction has parametrically smaller interaction constant, $J'^2/J
\ll J'$, than the SDW one, which diminishes its competitiveness even
further \cite{sato}.   {\em Hence, in the limit of weakly coupled
  chains, i.e. taking fixed intra-chain coupings and letting $J'
  \rightarrow 0^+$, the SDW always wins over the 2d SN.}

It is important to note here that sufficiently
close to the saturation our quasi-1d description, which assumes
linearly dispersing excitations, unavoidably breaks down. In the fully
polarized (saturated) phase, excitations are characterized by the
quadratic dispersion. The two-dimensional high-field nematic state
then occurs as a result of Bose-Einstein condensation (BEC) of magnon
pairs \cite{zhit}.  This represents a different order of limits: fixed
$J'>0$ (however small), and $M \rightarrow 1/2$.  Thus we expect a
wedge of SN phase intervening between the fully saturated state and
the nematic SDW, whose width approaches zero as $J' \rightarrow 0$.
We can further estimate the width versus $J'$ as follows.  In the
magnon description, the typical energy per unit chain length due to inter-chain pair magnon hopping is
proportional to $\Delta M J'^2/J $, where $\Delta M = \frac{1}{2}-M$,
while that due to magnon-magnon interactions across chains is $(\Delta
M)^2 J'$.  For small $\Delta M$ the former dominates, stabilizing the
pair magnon condensate, i.e. the SN, while for large $\Delta M$, the
latter is larger, provided $J'/J \ll 1$.  Equating the two, we obtain
$\Delta M_c \sim J'/J$, i.e. the SN-SDW boundary enters the 1d
saturation point linearly in the $J' - M$ plane.  This conclusion
agrees with other calculations\cite{sato}, which also argue the two-dimensional nematic state is
replaced by the two-dimensional longitudinal SDW state below 
critical magnetization $M_c = \frac{1}{2}-\Delta M_c \lesssim 1/2$.
The transition between the two phases is a first-order one, as is explained in Appendix~\ref{ap:rg}.

\subsection{2d nematic}
\label{sec:interch-coupl-2d}

In the SN phase, the low energy
properties are universal.  Then we can discuss them using whatever
technique is convenient.  We begin with an analysis using the coupled
chains description.  We can assume that $D_{\rm nem} < D_{\rm sdw}$,
or simply that we tune the SDW interaction to zero.  The results
should be physically applicable for $M_c<M<1/2$.

\subsubsection{chain mean field}
\label{sec:chain-mean-field}

Within the chain mean field approximation Eq.~\eqref{eq:nem-H} is then replaced by
\begin{equation}
 H_{\rm nem} \sim  \frac{J'^2}{J_1} \langle \cos[\beta \theta_y]\rangle \sum_y \int dx 
\cos[\beta \theta_y].
\label{eq:7}
\end{equation}

When the above expectation value is non-zero, there is long-range
nematic order.  What are the consequences on the level of single chain
description?  Obviously the mean field Hamiltonian is fully gapped,
the (-) modes being gapped already by Eq.~\eqref{eq:5} and the 1d
nematic modes by Eq.~\eqref{eq:7}.  Spin operators acting on the
ground state generate excitations above this gap.  

It is important to realize from the outset that the two sine-Gordon models,
represented by Eq.~\eqref{eq:5} for the (-) [specifically, $\varphi^{-}$] sector and Eq.~\eqref{eq:7} for the 
(+) [specifically, $\theta = \sqrt{2} \theta^{+}$] sector, are characterized by very different energy scales.
In the (-) sector the scale is set by the ferromagnetic chain exchange $J_1 \sim J_2$, while in the (+)
sector is it determined by a much smaller $J'^2/J_1 \ll J_{1,2}$. Correspondingly, soliton mass of the (-)
sector, which we denote as $\tilde{m}^{-}_s$, is much bigger than that for the (+) sector, denoted as $\tilde{m}^{+}_s$.
That is, $\tilde{m}^{-}_s \gg \tilde{m}^{+}_s$, as Figure~\ref{fig:nematic} shows. This important observation implies that 
lowest-energy excitations above the ground state of sine-Gordon models \eqref{eq:5} and \eqref{eq:7}
are given by the excitations of Eq.~\eqref{eq:7} alone, i.e. occur in the (+) sector. The (-) sector is much more massive
and in many respects can be treated as fully frozen.

Referring to Eq.~\eqref{eq:8}, one sees that $S^z$, being `dual' to $S^+$, generates solitons in
(+) sector. The solitons change corresponding topological charge $Q^{(+)}_{\rm charge} = \beta/(2\pi) \int dx ~\partial_x \theta$ by $\pm 1$.
At the simplest level we simply set $\varphi^{-}=0$ and no excitations are generated in this (-) sector, as discussed above.

In a full treatment, which we add here for completeness, one needs to allow for excitations of (-) modes as well.
We start by noting that nonlinear cosine term in \eqref{eq:5}
describes repulsive sector of the sine-Gordon model in which no breathers, which are soliton-antisoliton bound states,
are present. Formally this is easiest seen by calculating the corresponding parameter $\xi_{\rm nem}^{(-)} = (\sqrt{2}/R)^2/(8\pi - (\sqrt{2}/R)^2) = 1/(4\pi R^2 -1)$.
Since $4\pi R^2 \in (2, 1)$ for $M \in (0, 1/2)$, this dimensionless parameter $\xi_{\rm nem}^{(-)} \in (1, \infty)$. Now, 
according to \eqref{eq:m_n} (see also \cite{essler04}), the number of breathers is determined by the integer part of $1/\xi$,
which in the present case is zero. Thus, there are no breathers in the (-) sector. 
This consideration shows that excitations in the $\varphi^-$ sector are represented by the soliton-antisoliton continuum,
which starts above the threshold energy $2 \tilde{m}^{-}_s$, and thus costs considerable additional energy.

Therefore the spectrum of states created by $S^z$ begins with
a gapped but well-defined soliton mode at energy $\tilde{m}^{+}_s$ at
momentum $k_x = k_{\rm sdw}$.  At the
single-chain level, this is the minimum energy excitation in the  $S^z - S^z$ structure factor.
The next (in energy) mode corresponds to exciting soliton {\em together} with breather, and starts at energy 
$\tilde{m}^{+}_s + \tilde{m}^{+}_{n = 1,2}$, see \eqref{eq:nem-breathers} below.
Note that a different analysis is required to discuss the region
around $k_x=0$, as this region is controlled by a different, $\partial_x \varphi$, term in the
bosonization formulae. 

Considering transverse spin excitations, the situation is different.
Crucially, Eq.~\eqref{eq:8} shows that $S^\pm$ {\em always} generates solitons in the (-) sector, hence the 
minimal energy of the transverse mode excitation is  $\tilde{m}^{-}_s$, which, as discussed above, is quite large.
Since $S^\pm_y \sim e^{i \beta \theta_y/2}$ as far as the (+) sector is concerned, we expect that the nematic mean-field \eqref{eq:7}
results in the finite vacuum-to-vacuum matrix element $\langle \exp[i\beta \theta_y/2]\rangle \neq 0$.
This means that $S^\pm$ does not need to generate any excitation in the
(+) sector, or if it does, it generates a two-particle breather (or, solitons and antisolitons in equal numbers). 

The breathers of the  (+) sector have masses
\begin{equation}
\label{eq:nem-breathers}
\tilde{m}^{+}_n = 2 \tilde{m}^{+}_s \sin[\frac{\pi}{2} \xi_{\rm nem}^{(+)} ~n] ~\text{for}~ n = 1, 2, ...[1/\xi_{\rm nem}^{(+)}],
\end{equation}
where $\tilde{m}^{+}_s$ denotes the soliton mass of the sine-Gordon model Eq.\eqref{eq:7}.
Parameter $\xi_{\rm nem}^{(+)}$ here is given by
$\xi_{\rm nem}^{(+)} = (\sqrt{2}\beta)^2/(8\pi - (\sqrt{2}\beta)^2) = (1/(\pi R^2) -1)^{-1}$.
Hence, $\xi_{\rm nem}^{(+)} \in (1, 1/3)$ for $M \in (0, 1/2)$, so that $1/\xi_{\rm nem}^{(+)} \leq 3$ for $M \sim 1/2$,
resulting in two breathers present in the excitation spectrum of the model 
in the most relevant magnetization range outside the immediate vicinity of the saturation.

It is interesting to note that, according to Ref.\onlinecite{essler04}, {\em odd}-numbered (with mass $\tilde{m}^+_1$) and {\em even}-numbered 
(with mass $\tilde{m}^+_2$) breathers contribute differently to matrix elements of $\sin[\beta \theta/2]$ and $\cos[\beta \theta/2]$ operators.
Specifically, operators even under `charge conjugation' $\theta \to -\theta$, such as  $\cos[\beta \theta/2]$, couple the ground state to the even-numbered breathers only,
while the odd ones, such as $\sin[\beta \theta/2]$, connect only to the odd-numbered breathers. Since these two kinds of breathers are characterized by the
different masses, $\tilde{m}^+_1 < \tilde{m}^+_2$, transverse spin correlation functions $\langle S^x S^x \rangle$ and $\langle S^y S^y \rangle$ are characterized
by the different excitation gaps {\em above} the soliton mass energy $\tilde{m}^{-}_s$.
In other words, even though transverse spin correlations are short-ranged and disordered in the two-dimensional
nematic state, their {\em high-energy structure} is sensitive to the fact that the $U(1)$ symmetry is broken by Eq.\eqref{eq:7}, 
i.e. by the two-magnon condensation.

To summarize, on the chain mean-field level, two-dimensional nematic state is characterized
by massive excitations in both longitudinal and transverse channels. The minimal energies of these 
are $\tilde{m}^{+}_s$ and $\tilde{m}^{-}_s$, correspondingly. A rather large gap in the transverse structure factor, $\tilde{m}^{-}_s$,
appears already on the level of a single chain, as expected for a `bosonic'
superconductor such as high-field spin nematic here.

\subsubsection{Susceptibilities}
\label{sec:susceptibilities}

Turning to 2d susceptibilities now, we first consider the role of
collective modes.  The nematic state {\sl does} spontaneously break
the continuous U(1) rotation symmetry about the $z$ axis, so there
indeed must be a Goldstone mode.  It should appear as a {\em mode},
i.e. a pole with large spectral weight, however, only in the nematic
order parameter susceptibility, i.e. a four-spin correlation
function.   

We conclude that the Goldstone mode does {\em not} appear near
$k_x=k_{\rm sdw}$ in $\chi_{2d}^{zz}$ nor near $k_x=\pi$ in
$\chi_{2d}^{xy}$.  Thus we expect all excitations at these momenta to
remain gapped, just as they are in the single chain mean-field
treatment.  This is confirmed by an RPA treatment, which, due to the
lack of qualitative modifications of the single-chain behavior, we do
not present in detail here. For example, $\chi_{2d}^{zz}$ should have functional form
of equation \eqref{eq:2d-zz} but with replacement of the breather mass
$\Delta_1$ by the soliton mass of the model of Eq.~\eqref{eq:7}, and
similar replacement for the parameters $C_z$ and $Z_z$.  Obviously,
the RPA treatment will also restore transverse dispersion, which is
simply given by $J' \cos[k_y]$.

The one place where the nematic Goldstone mode must appear, on general
principles, albeit with small spectral weight, is
at low energy near $k=0$ in the correlation function of the conserved density which
generates the broken symmetry, which in this case is just the
longitudinal spin density $S^z_y(x)$.  This is rather tricky to
capture using bosonization and the RPA, so we instead obtain it from general principles.

Because this weight is a universal property of the two-dimensional
spin-nematic state, it can be obtained a phenomenological effective
field theory description.  The spin nematic can be regarded in this
sense as simply a condensate of bound pairs of spin flips - the quanta
of the $\psi$ field.  This is described by the usual action for a Bose
gas,
\begin{eqnarray}
S & &=  \int dx d\tau \sum_y \{ \frac{1}{2}(\psi_y \partial_\tau \psi_y^+
- \psi_y^+ \partial_\tau \psi_y) + \frac{1}{2m_x} \partial_x
\psi_y^+ \partial_x \psi_y \nonumber \\
&& - 
\mu |\psi_y|^2 + \frac{u}{2} |\psi_y|^2 |\psi_{y+1}|^2 - c J'^2 (\psi^+_y \psi_{y+1} + \text{h.c.}) \}
\label{action-2d-nem}
\end{eqnarray}
The $u$-term describes interchain interaction of longitudinal spin components, hence $u \sim J'$.
The last term arose from transverse hopping as in
Eq.~\eqref{eq:nem-H}.  It can be of course
written as $\frac{1}{2m_y} \partial_y \psi^+ \partial_y \psi$ at low
energies when the $\sum_y$ can be replaced by an integral. Introducing the
standard parameterization $\psi_y(x) = \sqrt{\rho(x,y)} e^{i \theta(x,y)}$ we obtain
\begin{eqnarray}
&&S = \int dx dy d\tau \Big( - i\rho \partial_\tau \theta + \frac{u}{2} (\rho-\rho_0)^2 +\\
&& \frac{1}{8\rho_0} [\frac{(\partial_x\rho)^2}{m_x} +  \frac{(\partial_y\rho)^2}{m_y}]  + \rho_0 [\frac{(\partial_x \theta)^2}{2m_x} + \frac{(\partial_y \theta)^2}{2m_y}]\Big). \nonumber
\end{eqnarray}
As usual, $\rho_0 = \mu/u$. The first term here shows that $\rho$ and
$\theta$ are a canonical pair, which implies the expected soft mode in
the density fluctuations.  Integrating out $\delta\rho = \rho -
\rho_0$ we obtain a canonical action for the Goldstone mode of the nematic
order parameter \eqref{eq:6}
\begin{equation}
S_\phi = \int dx dy d\tau \{ \rho_0 [\frac{(\partial_x \theta)^2}{2m_x} + \frac{(\partial_y \theta)^2}{2m_y}] + \frac{1}{2u} (\partial_\tau \theta)^2\}.
\end{equation}
This mode cannot be observed in $\langle S^+ S^-\rangle$ correlator.

If we instead integrate out the phase $\theta$, we obtain, switching
to the $(\omega, {\bf k})$ representation,
\begin{equation}
S_{\delta\rho}= \int d{\bf k} d\omega_n \{\frac{\omega_n^2}{4\rho_0 \epsilon_k}
+ \frac{\epsilon_k}{4\rho_0} + \frac{u}{2}\} \delta\rho_k \delta\rho_{-k} ,
\end{equation}
where $\epsilon_k = \frac{k_x^2}{2m_x} + \frac{k_y^2}{2m_y}$.
This action immediately translates into the following density-density
correlation function:
\begin{equation}
\langle \delta\rho_k \delta\rho_{-k}\rangle =
\frac{2\rho_0\epsilon_k}{\omega_n^2 + \epsilon_k (\epsilon_k + 2 \rho_0 u)}, 
\label{eq:rho}
\end{equation}
Analytically continuing Eq.~\eqref{eq:rho} to real frequency, we
obtain the spectral function, 
\begin{equation}
  \label{eq:25}
  {\text{Im}}\chi_{2d}^{zz}(k,\omega) \sim \frac{\rho_0 \epsilon_k}{\omega_B(k)}
  \delta(\omega - \omega_B(k)),
\end{equation}
where the frequency of the Bogoliubov mode is $\omega_B(k) = \sqrt{2 \rho_0
  u \epsilon_k + \epsilon_k^2}$.  At small momentum, the first term
in the square root dominates, $\omega_B(k) \sim
\sqrt{2\rho_0 \epsilon_k}$: the spectrum is linear (acoustic) and isotropic up to a
constant rescaling, i.e. $\omega_B(k) \propto
\sqrt{k_x^2/m_x + k_y^2/m_y}$, and the susceptibility becomes
\begin{equation}
  \label{eq:26}
 {\text{Im}}\chi_{2d}^{zz}(k,\omega) \sim \omega_B(k) \delta(\omega - \omega_B(k)),
\end{equation}
so that the weight of the acoustic Bogoliubov mode vanishes linearly
and isotropically with $k$.  A similar linear vanishing was observed
for the contribution of the phason in Eq.~\eqref{eq:27}, but in that
case the weight, though linear in $k$, was highly anisotropic.  The
difference between the isotropy found here and the anisotropy found
for the SDW originates from the physical difference that the nematic
represents a state of broken {\em internal} $U(1)$ symmetry,
unconnected with real (or momentum) space, while the SDW is a state of
broken translational symmetry, and hence the phason is intimately tied
to real and momentum space, influencing differently correlations along
or normal to the SDW wavevector.  

The
above density-density correlation function represents the physical
longitudinal spin-spin one, i.e. $\langle S^zS^z\rangle$, and hence is
observable in inelastic neutron scattering.
A similar observation has been made in Ref.~\cite{syro} by considering the
dynamic properties of the two-magnon condensate.  We note that the
general property that the spectral weight vanishes linearly in $k$
near $k=0$, shared by the nematic and the SDW, is required on general ground since the
total spin is conserved, and both the nematic and SDW states are
compressible, i.e. have finite non-zero susceptibility to a field
along the $z$ axis.

To summarize, the Goldstone model in the nematic case appears only in
the vicinity of the Brillouin zone center, and with small spectral
weight that vanishes as $k\to 0$.  By contrast, in the SDW state, the
``phason'' mode appears at the SDW wavevector, with {\em divergent}
spectral weight.  

\section{Discussion}
\label{sec:discussion}

\subsection{Discriminating SDW and SN phases}
\label{sec:discr-sdw-sn}

In this paper, we discussed the spectral properties of SDW and SN
phases, pointing out means to distinguish them.  At low energy, the
principle distinction is the phason mode, which gives power-law
spectral weight at ${\bf k}={\bf k}_{SDW}$ in the SDW state, which is
not present in the SN.  The other spectral distinctions were reviewed already
in the Introduction and Figs.~1-2, so will not be discussed further
here.

There are other ways to differentiate the SN and SDW, however.  One is
through their static order.  The SN has really no observable static
order in the spin structure factor.  By contrast, the SDW has static
order of the longitudinal $S^z_i$ moments.  This is clearly an
observable difference.  

In thinking about the SDW order, it is important to consider
the effects of quenched disorder.  The broken symmetry of the SDW
state is in fact just translational symmetry.  Hence, any defects act
as random fields on the SDW order parameter -- i.e. collective
pinning (see for example Ref.\cite{PhysRevB.17.535} in the context of CDWs).  It is well established that pinning of this type inevitably
destroys the long-range order of the SDW state (some exotic
``Bragg glass'' order\cite{giamarchi1995elastic} may survive as a distinct phase, though this is
not proven).  Consequently, a peak with finite correlation length should be
observed at the SDW wavevector in the $S^z-S^z$ structure factor in
the SDW state.  Furthermore, pinning will modify the thermal
transition from the paramagnetic to SDW state, which in the absence of
disorder would be expected to be XY-like.  The specific heat
singularity of the XY transition will be reduced and rounded.  

By contrast, the SN state breaks the {\em internal} spin-rotation
symmetry, and thus is not strongly effected by disorder.  In a
Heisenberg model, it would be expected to display an thermal XY
transition which unlike for the SDW is not rounded by disorder.
However, we should note that typically there will be some spin-orbit
coupling effects such as Dzyaloshinskii-Moriya interactions or
symmetric exchange anisotropy that anyway remove the continuous
rotation symmetry of the Heisenberg model about the field axis.  In
that case, the symmetry may be reduced to a discrete one, or none at
all.  This will certainly modify the SN transition, either to a
discrete universality class such as Ising (which has a stronger
specific heat singularity), or remove it entirely (if there is
insufficient rotation symmetry, then the SN order becomes no longer
spontaneous).

\subsection{Experiments}
\label{sec:experiments}

The list of materials realizing SDW and/or SN phases is pretty short.

The spin-1/2 Ising-like antiferromagnet BaCo$_2$V$_2$O$_8$ was, to our
knowledge, the first insulating material to realize collinear SDW
order, along the lines of the scenario outlined in
Section~\ref{sec:ising-anisotropy}.  Experimental confirmations of
this include specific heat \cite{Kimura2008} and neutron diffraction
\cite{Kimura2008a} measurements.  The latter one is particularly
important as it proves the linear scaling of the SDW ordering wave
vector with the magnetization, $k_{\rm sdw} = \pi(1-2M)$, predicted in
\cite{Okunishi2007}.  Subsequent NMR \cite{Klanjsek2012}, ultrasound
\cite{Yamaguchi2011} and neutron scattering \cite{Canevet2013}
experiments have refined the phase diagram and even proposed the
existence of two different SDW phases \cite{Klanjsek2012} stabilized
by competing interchain interactions.

Most recently, the spin-1/2 magnetic insulator LiCuVO$_4$ has emerged
\cite{Enderle2005,Nishimoto2012} as a promising candidate to realize
both a high-field spin nematic phase, in a narrow region below the
(two-magnon) saturation field (which is about $45$ T), as well as an
incommensurate collinear SDW phase at lower fields (which occupies a
huge magnetization/field interval, extending down to about $7.5$
T). In fact, the material seems to nicely realize the theoretical
scenario outlined in Section~\ref{sec:spin-nematic-chains}: despite
being in a one-dimensional spin-nematic state
\cite{Kolezhuk2005,McCulloch2008}, the chains order into a
two-dimensional nematic phase only in the immediate vicinity of the
saturation field \cite{Svistov2011}. At fields below that narrow
interval, which we estimated in Sec.~\ref{sec:comp-betw-2d} to be of
the order of $\Delta M_c \sim J'/J$, the ordering is instead into an
incommensurate longitudinal SDW state. Evidence for the latter
includes detailed studies of NMR line shape
\cite{Buttgen2007,Buttgen2010,Buttgen2012,Nawa2013}, which
convincingly exclude spin ordering transverse to the field, and
neutron scattering \cite{Masuda2011,Mourigal2012} studies. The neutron
scattering observes linear scaling of the SDW ordering momentum with
magnetization \cite{Masuda2011}. Using polarized neutrons,
Ref.\cite{Mourigal2012} has established non-spin-flip character of the
elastic neutron scattering (and the absence of spin-flip scattering)
at magnetic field above approximately $10$ Tesla, which strongly
points to the development of $U(1)$-preserving 2d SDW order.
(The low-field phase of the material, which is characterized by a more conventional 
vector chiral order, can be explained by a moderate easy-plane anisotropy 
of the exchange interaction \cite{Heidrich-Meisner2009}).
It should be noted that the authors of Ref.\onlinecite{Mourigal2012} interpret
their findings in terms of nematic bond order, which, in our opinion,
is not realized for intermediate magnetization values within the simple
model of weakly coupled spin nematic chains.  Indeed, the observations
of finite correlation length and rounded specific heat singularity in
their paper are very much in accord for the expectations in a pinned
SDW state, as discussed in the previous subsection.  It would be very
interesting to search for the predicted linear phason mode with the
help of inelastic neutron scattering.

Last, but not least, are spin-1/2 triangular lattice antiferromagnets
Cs$_2$CuCl$_4$ and Cs$_2$CuBr$_4$, whose geometric structure of which
is rather close to the third model, of
Section~\ref{sec:spat-anis-triang}, considered in this paper.  The
first of these unfortunately appears to be strongly disturbed by the
weak (of the order of several percent) residual inter-plane and
Dzyaloshinskii-Moriya (DM) interactions which dominate the
magnetization process \cite{extreme} and produce in a complex and
highly anisotropic h-T phase diagram \cite{Tokiwa2006}.  However, it is
worth mentioning that this was perhaps the first spin-1/2 material
studied for which an SDW-like ordering wave vector,
scaling linearly with magnetic field in an about 1 Tesla wide
interval (denoted as phase `S' in \cite{Coldea2001}), was observed in neutron scattering studies.

The magnetic response of Cs$_2$CuBr$_4$ is quite different and includes a
prominent commensurate longitudinal phase: the up-up-down
magnetization plateau at $M=M_{\rm sat}/3$
\cite{Ono2003,Ono2005,Fujii2007}. As discussed extensively in
\cite{extreme,chen2013}, in the limit of weak interchain interaction
$J'\ll J$, the magnetization plateau phase can be understood as a
commensurate version of the incommensurate longitudinal SDW phase (see
also Appendix~\ref{ap:plateau}). This connection makes it plausible
that an SDW phase may be `hiding' in the complex phase diagram of
Cs$_2$CuBr$_4$ \cite{Fortune2009}, though the estimates of $J'/J$ are
not so small.  An inelastic neutron scattering study of the gapped
phason at $M=M_{\rm sat}/3$ magnetization plateau, as well as that of
gapped transverse spin excitations, could reveal the
nature of this interesting frustrated antiferromagnet.

We hope that our work will stimulate further studies of the unusual ordered 
phases of frustrated  low-dimensional quantum magnets.

\begin{acknowledgements}
We would like to thank C. Broholm, R. Coldea, F. Essler, A. Furusaki, E. Fradkin, E. Mishchenko, M. Mourigal, L. Svistov, and M. Takigawa for useful discussions.
We especially thank F. Essler for pointing Ref.~\onlinecite{destri} to
us. This work is supported by NSF grant DMR-12-06809 (LB) and NSF DMR-12-06774 (OAS).
\end{acknowledgements}

\appendix
\section{Correcting sine-Gordon}
\label{ap:sG}

Here we describe how to correct sine-Gordon ground state energy. We start with Bethe ansatz result for the energy of the lattice model,
as given by eq.2.69 of Ref.\onlinecite{destri}:
\begin{equation}
e_{0}(a) = \frac{2}{a^2} \int_0^\infty \frac{dt}{t} \frac{\sin[4 \theta t]}{\cosh[\gamma t]} \frac{\sinh[(\pi - \gamma)t]}{\sinh[\pi t]} .
\label{eq:ap1}
\end{equation}
Here $\theta$ and short-distance cut-off $a$ determine soliton mass $m_s$ via
\begin{equation}
m_s = \frac{4}{a} e^{-\pi \theta/\gamma},
\label{eq:ap2}
\end{equation}
while $\gamma = \pi/(1+\xi)$ as can be checked later by comparing the final result with other tabulated forms.
The continuous limit corresponds to $a\to 0$ while $\theta\to \infty$ so that $m_s$ stays constant.

The idea is to solve \eqref{eq:ap1} and take the continuous limit, and drop everything that disappears when $a\to 0$. 
Because of $a^{-2}$ factor in front of \eqref{eq:ap1} it seems clear that result should be proportional to $m_s^2$, but let's see.

Introduce contour integral
\begin{equation}
I = \int _C f(t) \equiv \int _C \frac{dt}{t} \frac{e^{i 4 \theta t}}{\cosh[\gamma t]} \frac{\sinh[(\pi - \gamma)t]}{\sinh[\pi t]} 
\label{eq:ap1}
\end{equation}
where C is the contour $C = (-\infty,-\epsilon) \bigcap C_\epsilon \bigcap (\epsilon, \infty) \bigcap C_R$, where $C_\epsilon$ goes over the origin
from above (and $\epsilon \to 0$ of course) in clockwise fashion while $C_R$ is the standard large semi-circle traveled counterclockwise 
in the upper $\text{Im}[t] > 0$ half-plane, with $R\to \infty$. Doing residues and everything else we find 
\begin{equation}
a^2 e_0(a)/2 = \frac{\pi - \gamma}{2} + \pi \sum {\text{Res}}[f(t)] .
\end{equation}
The first term comes from $C_\epsilon$. The residues of $f(t)$ are of two kinds: from $\sinh[\pi t]=0$ we get $t_n = i n$, where $n=1,2,3...$,
while $\cosh[\gamma t]=0$ produces $t_k = i (k - 1/2) \pi/\gamma$, with  $k=1,2,3...$.

Thus 
\begin{eqnarray}
a^2 e_0(a)/2 &=& \frac{\pi - \gamma}{2} - \pi \sum_{n=1} \frac{e^{-4\theta n}}{\pi n} \tan[\gamma n] + \\
&&+\pi \sum_{k=1} \frac{e^{-4\pi \theta (k-1/2)/\gamma}}{\pi (k-1/2)} \cot[\frac{\pi^2}{\gamma}(k-\frac{1}{2})] \nonumber
\end{eqnarray}

We observe that the standard result
\begin{equation}
e_0(0) = m_s^2  \cot[\frac{\pi^2}{2\gamma}] = - m_s^2 \tan[\frac{\pi \xi}{2}]
\end{equation}
is obtained from $k=1$ contribution from the last term. Everything else scales as higher than second power of $m_s a$ and
disappears in the $a\to 0$ limit.

Note however that at $\gamma=\pi/2$ the soliton mass \eqref{eq:ap2} $m_s \sim e^{-2 \theta}$, so that 
the first member of the first sum, $n=1$, too scales as $e^{-4\theta} \sim (m_s a)^2$, and thus must be kept.
That is, at $\gamma=\pi/2$ the two poles merge. We then obtain
\begin{eqnarray}
a^2 e_0(a)/2 &=& \frac{\pi - \gamma}{2} + \big(\frac{m_s a}{4}\big)^2 \big\{2 \cot[\frac{\pi^2}{2\gamma}] - \\
&&\big(\frac{m_s a}{4}\big)^{\frac{4\gamma}{\pi}-2} \tan[\gamma]\big\} + O((m_s a)^{p > 2}) \nonumber
\label{eq:ap3}
\end{eqnarray}
Taking the limit $\gamma\to \pi/2$, we immediately obtain finite result for the ground state energy density
\begin{equation}
e_0(a) = \frac{\pi}{2a^2} + \frac{m_s^2}{4\pi} \ln[\frac{m_s^2 a^2}{16 e}]
\label{eq:ap4}
\end{equation}

Next we need to realize that $\gamma = \pi/2$ ($\xi = 1$) corresponds to the non-interacting Thirring model, see for example \cite{thirring},
\begin{equation}
H_0 = \sum_k u k (a^+_{1k} a_{1k} - a^+_{2k} a_{2k}) + m_0 (a^+_{1k} a_{2k} + a^+_{2k} a_{1k})
\end{equation}
spectrum of which is given by massive fermions with dispersion $\pm \sqrt{u^2 k^2 + m_0^2}$. The ground state energy is found as
(all negative levels are filled) 
\begin{eqnarray}
E_{\rm Thirring} &=& -\int_{-\Lambda}^\Lambda \frac{dk}{2\pi} \sqrt{u^2 k^2 + m_0^2} = \nonumber\\
&&= - u \frac{\Lambda^2}{2\pi} + 
\frac{m_0^2}{4\pi u}\ln[\frac{m_0^2}{4u^2 \Lambda^2}]
\label{eq:ap5}
\end{eqnarray}
Clearly it matches, in its scaling (mass-dependent) part, Eq.\eqref{eq:ap4}. Since the field-theory expression is written in dimensionless units, we
can identify $m_s = m_0/u$ and $a = 2\sqrt{e}/\Lambda$. Taking $\Lambda = \pi$ in \eqref{eq:ap5} suggests $a=1.05$.

All of this shows that the free energy density of the sine-Gordon model should be modified to 
\begin{equation}
F_{\rm new} = - \frac{m_s^2}{8}\Big( 2 \tan[\frac{\pi \xi}{2}] + \Big(\frac{m_s a}{4}\Big)^{2(1-\xi)/(1+\xi)} \tan[\frac{\pi}{1+\xi}]\Big)
\label{eq:ap6}
\end{equation}
For $\xi < 1$ the second term is subleading correction which, at $\xi=1$, serves to cancel unphysical divergence of the first term. 

Next, we apply the obtained result to the self-consistent solution of the chain mean-field. 
As before, $\overline\Phi = -(1/2) \partial F_{\rm new}/\partial \mu = -(1/2) (\partial F_{\rm new}/\partial m_s^2) (\partial m_s^2/\partial \mu)$.
Using
\begin{equation}
\frac{d m_s^2}{d\mu} = \Big(\frac{2 \Gamma(\xi/2)}{\sqrt{\pi} \Gamma((1+\xi)/2)}\Big)^2 \Big(\frac{\pi \Gamma(1/(1+\xi))}{\Gamma(\xi/(1+\xi))}\Big)^{1+\xi}
(1+\xi) \mu^\xi,
\end{equation}
which is obtained from \eqref{eq:mu-m0}, we can solve for $\mu = \gamma_{\rm sdw} \overline\Phi$:
\begin{eqnarray}
\label{eq:ap7}
&&(\mu/v)^{1-\xi} = \frac{1+\xi}{8} \tan[\pi \xi/2] A_1^2 A_2^{1+\xi} (\gamma_{\rm sdw}/v) \\
&&\times\Big(1 - \frac{1}{8}\tan[\pi/(1+\xi)] A_1^\frac{4}{(1+\xi)} A_2^2 ~Q^\frac{(1-\xi)}{(1+\xi)} (\gamma_{\rm sdw}/v)\Big)^{-1}. \nonumber
\end{eqnarray}
Here $Q = \frac{a^2}{16}$, $A_1 = \frac{2 \Gamma(\xi/2)}{\sqrt{\pi} \Gamma((1+\xi)/2)}$, $A_2 = \frac{\pi \Gamma(1/(1+\xi))}{\Gamma(\xi/(1+\xi))}$.
Notice that the whole denominator in \eqref{eq:ap7} is the result of the new (second) term in \eqref{eq:ap6}. Both tangents diverge at $\xi=1$, but their ratio is finite,
and the right-hand-side goes to 1 in this limit. 

Once \eqref{eq:ap7} is solved, the soliton mass is found as
\begin{equation}
m_s = v A_1 \Big(\frac{\mu}{v} A_2\Big)^{(1+\xi)/2}
\label{eq:ap8}
\end{equation}
This equation is plotted in Fig.~\ref{fig:delta} for the particular case of spatially anisotropic triangular lattice model with $\gamma_{\rm sdw} = J' A_1^2 \sin[\pi M]$.

\section{Alternative derivation of the phason mode}
\label{ap:phason}

Here we present an alternative, Ginzburg-Landau action derivation of the phason mode and its dispersion in the 2d collinear SDW state.
We start with the partition function of $\varphi_y(x)$ field
\begin{eqnarray}
&&Z_{\rm sdw} = \int D\varphi \exp\{ - A_0 + \sum_y \int d\tau dx \gamma_{\rm sdw} \nonumber\\ 
\times&& \cos[2\pi(\varphi_y - \varphi_{y+1})/\beta] \} = \nonumber\\
&& = \int D\varphi \exp\{ - A_0 + \int (d{\bf k}) J'_{\rm zz}(k_y) \vec{\sigma}_{\bf k} \cdot \vec{\sigma}_{\bf -k}\},
\label{eq:ap9}
\end{eqnarray}
where $A_0 = \sum_y \int d\tau dx \frac{1}{2}\{ (\frac{1}{v} \partial_\tau \varphi_y)^2 + v (\partial_x \varphi_y)^2 \}$ is the action of decoupled chains,
inter-chain interaction $J'_{\rm zz}(k_y) = \gamma_{\rm sdw} \cos[k_y]$ is the same as in Section~\ref{sec:rpa}, and 
$\vec{\sigma}(x,y) = (\cos[2\pi \varphi_y(x)/\beta], \sin[2\pi \varphi_y(x)/\beta])$ stands for a SDW vector, 
and $\vec{\sigma}_{\bf k}$ is its Fourier transform. Finally, $\int (d{\bf k}) \equiv \int d\omega d k_x d k_y/(2\pi)^3$.

We next apply Hubbard-Stratanovich identity to decouple interchain cosine term with the help of the vector field $\vec{\Psi}_y(x,\tau)$,
\begin{eqnarray}
\label{eq:ap10}
&&Z_{\rm sdw} = \int D\varphi D\vec{\Psi} \exp\{ - A_0 + \\
&&+ \int (d{\bf k}) [\frac{1}{4 \gamma_{\rm sdw}}(1 + \frac{1}{2} k_y^2) \vec{\Psi}_{\bf k} \cdot  \vec{\Psi}_{\bf -k} + \vec{\Psi}_{\bf k} \cdot \vec{\sigma}_{\bf -k}]\}, \nonumber
\end{eqnarray}

Inside SDW phase $\vec{\Psi}_y(x,\tau)$ takes on finite expectation value,  $\langle |\vec{\Psi}_y(x,\tau)|\rangle = \rho \neq 0$ 
and consequently we parameterize it as $\vec{\Psi}_y(x) = \rho (\cos[2\pi \Phi(x,y)/\beta], \sin[2\pi \Phi(x,y)/\beta])$ and treat the magnitude of
the order parameter $\rho$ as a constant. Also note that in \eqref{eq:ap10} we have expanded $\cos[k_y]$ in $J'_{\rm zz}(k_y)$ about the minimum at $k_y=0$.
We then observe that in continuum approximation
$\int (d{\bf k}) (1 + \frac{1}{2} k_y^2) \vec{\Psi}_{\bf k} \cdot  \vec{\Psi}_{\bf -k} = \rho^2 \int d\tau dx dy \{ 1 + \frac{1}{2} (\partial_y \Phi)^2\}$, while
$\int (d{\bf k}) \vec{\Psi}_{\bf k} \cdot \vec{\sigma}_{\bf -k} = \rho \int d\tau dx dy \cos[2\pi ( \Phi(x,y) - \varphi_y(x))/\beta]$. 

We now absorb phase $\Phi(x,y)$ into $\varphi_y(x)$ via the shift 
\begin{equation}
\varphi_y(x) = \tilde{\varphi}_y(x) + \Phi(x, y) .
\label{eq:ap11}
\end{equation}
This simple transformation changes $\cos[2\pi ( \Phi(x,y) - \varphi_y(x))/\beta]$ into the cosine term of the $2+1$-dimensional sine-Gordon model, 
$\cos[2\pi  \tilde{\varphi}_y(x)/\beta]$, which strongly pins $\tilde{\varphi}_y(x)$ to one of its minima.

As a result, \eqref{eq:ap10} can be re-written as
\begin{eqnarray}
&&Z_{\rm sdw} = \int D\tilde{\varphi} D\Phi \exp\{ - \frac{1}{2} \int d\tau dx dy [\frac{1}{v} (\partial_\tau \Phi)^2 + \nonumber\\
&& +v (\partial_x \Phi)^2 + \frac{\rho^2}{\gamma_{\rm sdw}} (\partial_y \Phi)^2 + \rho \cos[2\pi  \tilde{\varphi}_y(x)/\beta] + ...]\}
\label{eq:ap12}
\end{eqnarray}
Observe that in this expression $\rho$ plays the role of the pinning potential and provides $\tilde{\varphi}$ with a finite mass.
Correspondingly, the coupling between $\Phi$ and $\tilde{\varphi}$ fields, which is included in the omitted ``..." terms,
is irrelevant for energies/momenta much smaller than $\rho$. For example, the coupling such as $\partial_x \tilde{\varphi} \partial_x \Phi$ 
can be easily shown to only generate quartic (in derivatives or momenta) corrections, such as $\rho^{-1} (\partial_x^2 \Phi)^2$, to the leading
quadratic terms in \eqref{eq:ap12}.

Omitting such terms we observe that \eqref{eq:ap12} predicts linearly-dispersing phason mode $\Phi$ with dispersion
\begin{equation}
\omega^2 = v^2 k_x^2 + \frac{v \rho^2}{\gamma_{\rm sdw}} k_y^2.
\label{eq:ap13}
\end{equation}

It remains to relate $\rho = \langle |\vec{\Psi}_y(x,\tau)|\rangle$ to the SDW order parameter $\tilde\psi$ in Section~\ref{sec:sine-gordon}.
This is done via the following simple consideration: imagine adding source term $\sum_y \int d\tau dx ~\vec{\lambda} \cdot \vec{\sigma}$ to \eqref{eq:ap9}.
Upon Hubbard-Stratonovich decoupling in \eqref{eq:ap10} it is seen that $\vec{\lambda}$ couples to $\vec{\sigma}$ in the same way as
$\vec{\Psi}$ does. Hence the shift $\vec{\Psi} \to \vec{\Psi} - \vec{\lambda}$ removes the linear $\vec{\lambda} \cdot \vec{\sigma}$ term
simultaneously generating quadratic $(4 \gamma_{\rm sdw})^{-1} \int (\vec{\Psi} - \vec{\lambda})\cdot (\vec{\Psi} - \vec{\lambda})$ term.

On the other hand, 
\begin{eqnarray}
\tilde{\psi} &=& \langle \vec{\sigma}\rangle = Z_{\rm sdw}^{-1} \frac{\partial Z_{\rm sdw}}{\partial \vec{\lambda}}|_{\lambda =0} \\
&& = \frac{1}{2\gamma_{\rm sdw}} \langle \vec{\Psi}\rangle_{\lambda =0} \sim \frac{\rho}{\gamma_{\rm sdw}}.
\end{eqnarray}
Hence transverse velocity in \eqref{eq:ap13} can be estimated as $v_\perp^2  =  v \rho^2/\gamma_{\rm sdw} \sim v \gamma_{\rm sdw} (\tilde{\psi})^2$.
Since from \eqref{eq:ap7} $\mu/v \sim (\gamma_{\rm sdw}/v)^{1/(1-\xi)}$ and from Section~\ref{sec:sine-gordon} $\mu = \gamma_{\rm sdw} \tilde{\psi}$,
we find that $\tilde{\psi} \sim (\gamma_{\rm sdw}/v)^{\xi/(1-\xi)}$ and finally obtain
\begin{equation}
v_\perp^2 \sim v^2 (\gamma_{\rm sdw}/v)^{(1 + \xi)/(1-\xi)} = v^2 (\gamma_{\rm sdw}/v)^{4\pi R^2/(4\pi R^2 - 1)} ,
\label{ap:v-perp}
\end{equation} 
which results in the same scaling $v_\perp \sim J (J'/J)^{2\pi R^2/(4\pi R^2 -1)}$ as previously obtained in Section~\ref{sec:sine-gordon},
see in-line equation below \eqref{eq:disp-phason}, by insisting on the gaplessness of the longitudinal spin fluctuations.
The present consideration shows that the phason is indeed direct consequence of the formation of the 2d SDW order.

\section{Magnetization plateau}
\label{ap:plateau}

Approach developed in the previous Appendix~\ref{ap:phason} 
also explains the appearance of the magnetization plateaux inside the established SDW state. 
For this we need to go back to \eqref{eq:ap9} and allow for the nominally irrelevant terms
to be retained in the single chain action $A_0$. Such subleading terms still have to respect the symmetries of the two-dimensional
lattice. For the case of spatially anisotropic triangular lattice the required symmetry analysis was performed in Ref.\onlinecite{extreme}, Section III.D.

For convenience we briefly summarize it here. Inside the SDW phase, magnetization plateaux are possible when the ordering 
momentum of the SDW state $\pi(1-2M)$ is the rational fraction of the reciprocal lattice momentum $2\pi$, $\pi(1-2M) k = 2\pi \nu$, with integer $k$ and $\nu$.
This leads to the following allowed magnetization values
\begin{equation}
M^{(k, \nu)} = \frac{1}{2} (1 - \frac{2\nu}{k}).
\label{ap:plat1}
\end{equation}
Importantly, the integers $\nu$ and $k$ must satisfy the {\em same parity} constraint\cite{extreme}: $\nu$ must be of the same parity as $k$
(both are even or odd). Given this, the following $k$-th order umklapp term can be added to the SDW Hamiltonian (and, consequently, to the action
in \eqref{eq:ap9}):
\begin{equation}
H_{\rm umk}^{(k)} =  \sum_y \int dx ~t_k \cos[\frac{2\pi k}{\beta} \varphi_y(x)]
\label{ap:plat2}
\end{equation}
The amplitude of this $k$-th order umklapp can be estimated \cite{extreme} to scale as $t_k \sim J (J'/J)^{k^2/(8\pi R^2 -2)}$, where
the comptification radius $R$ depends on the magnetization $M^{(k, \nu)} $.

The strongest plateau is 1/3 magnetization plateau when $k=3, \nu=1$ and $M^{(3,1)} = \frac{1}{3} \times \frac{1}{2} = \frac{1}{6}$.
Note that SDW ground state is a necessary condition for the plateau existence - this tends to remove many of potential higher-order
plateaux, $k > 3$, as they require higher magnetization \eqref{ap:plat1}.

The effect of adding $t_k \cos[\frac{2\pi k}{\beta} \varphi_y(x)]$ to \eqref{eq:ap9} is easy to track - being of single-chain origin, it does not
affect steps leading to \eqref{eq:ap12}. Obviously substitution \eqref{eq:ap11} changes it into $t_k \cos[\frac{2\pi k}{\beta} \{ \tilde{\varphi}_y(x) + \Phi(x, y)\}]$.
This simple result has a very profound meaning: since $\tilde{\varphi}_y(x)$ is already pinned by the SDW 
potential $\rho \cos[2\pi  \tilde{\varphi}_y(x)/\beta]$ in \eqref{eq:ap12},
the added umklapp term simply becomes a {\em pinning potential for the phason} field $\Phi(x, y)$. 

Note that at this stage we are dealing with a two-dimensional sine-Gordon model which can be analyzed classically, see Refs. \cite{extreme,chen2013}.
It follows that once the commensurability condition is satisfied, $\Phi(x, y)$ is pinned and phason mode becomes gapped.
Its lowest energy excitations are given by kinks, which interpolate between degenerate minima of $\Phi$, and cost finite energy 
$\Delta_{\rm plat}\sim\sqrt{v \tilde{t}_k}$.

All of this allows us to generalize expression longitudinal susceptibility of the SDW \eqref{eq:2d-phason} to the two-dimensional
plateau state
\begin{equation}
  \chi_{\rm 2d; plat}^{zz}(q,\pi+q_y,\omega) \sim \frac{Z_{\rm zz;2d}}{\Delta_{\rm plat}^2 + (v^2 q^2 + v_\perp^2 q_y^2) - \omega^2} .
\label{eq:2d-phason-plateau}
\end{equation}
Here $q$ is measured from the commensurate with the lattice SDW momentum $k_{\rm sdw}^{(k,\nu)} = \pi (1 - 2 M^{(k,\nu)})$.
Refs.\onlinecite{extreme,chen2013} show that the plateau-SDW transition, driven by the sufficient deviation of the magnetic field away from
the `commensurate' value corresponding to \eqref{ap:plat1}, is of the commensurate-incommensurate (CIT) kind.

It should be clear that transverse spin fluctuations \eqref{eq:chi-perp} are not affected by the development of the plateau state and remain gapped 
as before.

\section{RG analysis of the SDW-SN transition}
\label{ap:rg}

To describe the competition between the SN and SDW phases near the
saturation field (low magnon density), we start with the boson action for
the pair magnon field $\psi_y(\tau,x)$
\begin{eqnarray}
S & &=  \sum_y \int dx d\tau  \{ \psi_y^+ \partial_\tau \psi_y
 + \frac{1}{2m_x} \partial_x
\psi_y^+ \partial_x \psi_y \nonumber \\
&& - t (\psi^+_y \psi_{y+1} + \psi^+_{y+1} \psi_{y})   - \mu |\psi_y|^2\nonumber\\
&&+ \frac{w}{2} |\psi_y|^4 +  \frac{u}{2} |\psi_y|^2 |\psi_{y+1}|^2 \}.
\label{ap:sdw-sn}
\end{eqnarray}
Here, in comparison with \eqref{action-2d-nem}, we have denoted $t= c
(J')^2/J_1$ and also included the in-chain repulsion $w$.  Because the
transition occurs at zero boson density, the parameter $w$, though
important for describing the interactions between bosons in the
unsaturated phase, does not affect scaling exponents, as we comment
further on below. This is all we
will need in order to consider the competition between the
density-density ($S^z-S^z$) interaction $u \sim J'$ {\em vs} and the
pair-tunneling $t \sim (J')^2/J_1$.

Denoting the spatial scale along the chain as $L$, we conclude that
$\tau \sim L^2$, so that the dynamical critical exponent $z=2$, as is
evident from the first line of \eqref{ap:sdw-sn}. Demanding that
in-chain kinetic energy is marginal, we observe that the field $\psi_y$
scales as $\psi_y \sim 1/\sqrt{L}$. This means that magnon density
$\Delta M = 1/2 - M = 2 \psi_y^+ \psi_y \sim 1/L$.

We can now consider how the various interactions renormalize.  We see
that the the two four-boson terms $u$ and $w$ are relevant and grow as
$L$, while the hopping $t$ (and chemical potential $\mu$) is more
relevant and grows as  $L^2$.  The growth of $w$ is well-understood:
for a single chain it implies that the magnons behave as hard-core
particles and indeed their density-fluctuations becomes those of free
fermions.  Because those fiducial fermions are free, this physics does
not modify the scaling dimensions of the density operators
$|\psi_y|^2$ and the growth of $u$ is not modified by this effect.
Furthermore, since the hopping $t$ can be considered in the single
boson sector, interactions also cannot modify its scaling dimension.
Hence the growth of $w$ has no effect upon the renormalization of $u$,
$t$, and $\mu$.  

We should stop the scaling at the scale $L_M \sim 1/\Delta M$, determined by the magnon
density. At that scale we must compare the renormalized $u \to u/\Delta M$
with the renormalized $t \to t/(\Delta M)^2$. Equating the two
renormalized interactions gives us critical density $\Delta M_c \sim
t/u \sim J'/J_1 \ll 1$.  For $\Delta M \ll t/u$ (low magnon density) we have
$t/(\Delta M)^2 \gg u/\Delta M$ (this is the tunneling-dominated SN phase)
while in the opposite limit of `high' density $\Delta M \gg t/u$ (but
still $\Delta M \ll 1$) we have instead $u/\Delta M \gg t/(\Delta
M)^2$ (the repulsion dominated SDW phase).  Thus, on reducing the
magnetization $M$ from the saturated value $M_{\rm sat} =1/2$, the
system transitions from the fully polarized state into a spin-nematic
one, via the condensation of magnon pairs. The SN phase occupies the
narrow magnetization interval $\Delta M_c \sim t/u \sim J'/J_1$. For
$M \leq 1/2 - \Delta M_c$ the ground state is the (paired) longitudinal
SDW.  This conclusion is identical to the energy scaling argument
presented in the end of Section~\ref{sec:comp-betw-2d}.

The SN-SDW transition is most likely discontinuous, as can be
understood from realizing that \eqref{ap:sdw-sn} (and
\eqref{action-2d-nem}) is mathematically equivalent to the low-energy theory of the XXZ model
with a magnetic field along the easy axis. The model is Ising-like, with
$u \gg t$, and is actually the one described by \eqref{eq:ising}. The
SN-SDW transition is then a version of the spin-flop transition, which
is a first-order transition \cite{Holtschneider2007}.

Another possibility for the SN-SDW on general grounds is that there is an intermediate co-existence
phase. That phase can occur as a result of instability of the
Bogoliubov mode $\omega_B(k) = \sqrt{ \epsilon_k (\epsilon_k + 2
  \rho_0 u_k)}$ which may occur at some ${\bf k} \neq 0$ due to
k-dependence of the interaction $u_k$ (which is Fourier transform of
$u$-term in \eqref{action-2d-nem}). Such an instability describes
crystallization, {\em i.e.}  modulation of density $|\psi_y|^2$ with
coordinate.  However, we expect that a first order transition is most
likely.  

\bibliography{sdw.bib}

\end{document}